\documentclass[a4paper]{article}

\usepackage[reqno,centertags,intlimits]{amsmath}
\usepackage{amssymb,amsfonts}
\usepackage{cite,epsfig,graphics}

\usepackage{ASDS}

\begin{document}


  \title{Accelerated sources in de~Sitter spacetime
         and the insufficiency of retarded fields\thanks%
         {Published in Phys. Rev. D {\bf 64}, 124020 (2001)}
         }

  \author{Ji\v{r}\'{\i} Bi\v{c}\'{a}k\thanks{e-mail: \texttt{bicak@mbox.troja.mff.cuni.cz}}\ \  and
          Pavel Krtou\v{s}\thanks{e-mail: \texttt{Pavel.Krtous@mff.cuni.cz}}\\
    {\small Institute of Theoretical Physics,}\\
    {\small Faculty of Mathematics and Physics, Charles University,}\\
    {\small V Hole\v{s}ovi\v{c}k\'{a}ch 2,
            180 00 Prague 8, Czech Republic}
    }

  \date{January 25, 2002} 

  \maketitle

\begin{abstract}
The scalar and electromagnetic fields
produced by the geodesic and
uniformly accelerated discrete charges in de~Sitter spacetime
are construc\-ted by employing the conformal relation between
de~Sitter and Minkowski space.
Special attention is paid to
new effects arising in spacetimes which,
like de~Sitter space, have \emph{spacelike}
conformal infinities. Under
the presence of particle and event horizons,
purely retarded fields (appropriately defined)
become necessarily singular or even cannot be constructed
at the \vague{creation light cones} ---
future light cones
of the \vague{points} at which
the sources \vague{enter} the universe.
We construct smooth (outside the sources) fields involving both
retarded and advanced effects, and analyze the fields
in detail in case of (i) scalar monopoles,
(ii) electromagnetic monopoles,
and (iii) electromagnetic rigid and
geodesic dipoles.

\vspace{2ex}

\begin{flushright}
\end{flushright}
\end{abstract}

\section{Introduction}
\label{sc:intro}

The de~Sitter's 1917 solution of the vacuum Einstein equations
with a positive cosmological constant $\Lambda$, in which
freely moving test particles accelerate away from one another,
played a crucial role in the acceptance of expanding standard
cosmological models at the end of the 1920s
\cite{Bertottietal:book,Peebles:book}. It
reappeared as the basic arena in steady-state cosmology in the 1950s,
and it has been resurrected in cosmology again in the context of
inflationary theory since the 1980's \cite{Peebles:book}. de~Sitter
spacetime represents the \vague{asymptotic state}
of cosmological models with $\Lambda > 0$ \cite{Maeda:1989}.

Since de~Sitter space shares with Minkowski space the property of
being maximally symmetric but has a nonvanishing constant
positive curvature and nontrivial global properties, it has been
widely used in numerous works studying the effects of curvature
in quantum field theory and particle physics (see, e.g.,
Ref.~\cite{BirrellDavies:book} for references).
Recently, its counterpart with a
constant negative curvature, anti--de~Sitter space, has received
much attention again from quantum field and string theorists
(e.g. Ref.~\cite{DeWittHerger:2000}).

These three maximally symmetric spacetimes of constant curvature
also played a most important role in gaining many valuable insights
in mathematical relativity. For example, both the particle
(cosmological) horizons and the event horizons for geodesic
observers occur in de~Sitter spacetime, and the Cauchy horizons in
anti--de~Sitter space (e.g. Ref.~\cite{HawkingEllis:book}).
The existence of the past
event horizons of the worldlines of sources producing fields on
de~Sitter background is of crucial significance for the structure
of the fields.

The existence of the particle and event horizons is intimately
related to the fact that de~Sitter spacetime has, in contrast with
Minkowski spacetime, two \emph{spacelike} infinities --- past and
future --- at which all timelike and null worldlines start and end
\cite{HawkingEllis:book}. Since the pioneering work of Penrose
\cite{Penrose:1963,Penrose:1965} it has been
well known that Minkowski, de~Sitter, and anti--de~Sitter
spacetimes, being conformally flat, can be represented as
parts of the (conformally flat) Einstein static universe.
However, the causal structure of these three spaces is globally
very different. The causal character of the conformal boundary
$\scry$ to the physical spacetime
that represents the endpoints at
infinity reached by infinitely extended null geodesics, depends on
the sign of $\Lambda$. In Minkowski space, these are \emph{null}
hypersurfaces ---
future and past null infinity,
$\scry^+$ and $\scry^-$. In de~Sitter space, both
$\scry^+$ and $\scry^-$ are \emph{spacelike};
in anti--de~Sitter space the conformal
infinity $\scry$ is not the
disjoint union of two hypersurfaces, and it is timelike.

\begin{figure}
  \begin{center}
  \epsfig{file=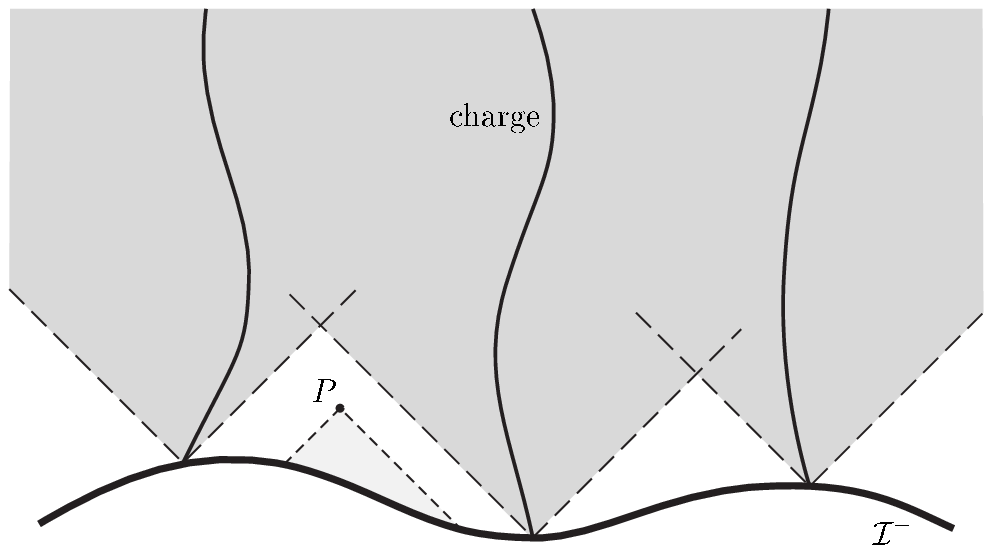}
  \end{center}
  \caption{Fields at spacelike $\scry^-$.}
  \figdescription{
  When past infinity $\scry^-$ is spacelike,
  and some discrete sources \vague{enter} the spacetime,
  then incoming fields must necessarily be
  present at $\scry^-$ and also at such points as $P$,
  the past null cones of which are not crossed
  by the worldlines of the sources.
  If this is not the case, inconsistencies arise.
  The past null cone of $P$ is shaded in light gray,
  whereas the future domain of influence
  of sources is shaded in dark gray.
  (Figure taken after Penrose \cite{Penrose:1964}.)
  }
  \label{fig:Penrose}
\end{figure}

Towards the end of his 1963 Les Houches
lectures \cite{Penrose:1964},
Penrose discusses briefly the zero rest-mass free fields with
spin $s$ in cosmological (not necessarily de~Sitter) backgrounds.
At a given point $P$, not too far from $\scry^-$, say, the field can be
expressed as an integral over quantities defined on the
intersection of the past null cone
of $P$ and $\scry^-$ (\vague{free
incoming radiation field}) plus contributions from sources whose
worldlines intersect the past null cone. However, the concept of
\vague{incoming radiation field} at
$\scry^-$ depends on the position of
$P$ if $\scry^-$ is spacelike
\cite{Penrose:1964,PenroseRindler:book}.
If there should be no incoming radiation at $\scry^-$
with respect to all \vague{origins $P$},
all components of the fields must vanish at $\scry^-$.
Imagine that spacelike $\scry^-$ is met by
the worldlines of discrete sources.
Then there will be points $P$ near $\scry^-$
whose past null cones
will not cross the worldlines --- see
Fig.~\ref{fig:Penrose}. The field
at $P$ should vanish if an incoming field is absent. This, however,
is not possible since the \vague{Coulomb-type} part of the field
of the sources cannot vanish there (as follows from Gauss's law).
Penrose \cite{Penrose:1964} thus
concludes that ``if there is a particle horizon, then purely
retarded fields of spin $s>\frac{1}{2}$ do not exist for general
source distributions.''%
\footnote{The corresponding result
  holds for spacelike $\scry^+$ and advanced fields.}
(The restriction on $s$ follows from the number of
arbitrarily specifiable initial data for
the field with spin $s$ --- see Ref.~\cite{Penrose:1964}.)
Penrose also emphasized that the result
depends on the definition
of advanced and retarded fields, and
``the application of the result to actual
physical models is not at all clear cut ... .'' This
observation was reported in a somewhat more detail at the
meeting on ``the nature of time'' \cite{Penrose:1967},
with an appended discussion (in which, among others,
Bondi, Feynman, and Wheeler participated)
but technically it was not developed futher since 1963.
In a much later monograph Penrose
and Rindler \cite{PenroseRindler:book} discuss
(see p.~363 in Vol.~II) the fact that
the radiation field is \vague{less invariantly} defined when
$\scry$ is spacelike than when it is null, but no comments or
references are given there on
the absence of \vague{purely retarded fields.}

One of the purposes of this paper is to study
the properties of fields
of pointlike sources \vague{entering} the de~Sitter universe across
spacelike $\scry^-$. We thus provide a specific physical model on
which Penrose's observation can be demonstrated and analyzed. We
assume the sources and their fields
to be weak enough so that they
do not change the de~Sitter background.

In de~Sitter space we identify retarded (advanced) fields of
a source as those which are in general nonvanishing only in the
future (past) domain of influence of the source.
As a consequence, purely retarded
(advanced) fields have to vanish
at the past (future) infinity. Adopting this definition we shall
see that indeed purely retarded fields produced by pointlike
sources cannot be smooth or even do not exist. We
find this general conclusion to be true not only for charges
(monopoles and dipoles) producing
electromagnetic fields $(s = 1)$
but, to some degree, also for
scalar fields $(s = 0)$ produced by scalar charges.

In general, purely retarded fields of monopoles and dipoles
become singular on the past horizons
(\vague{creation light cones}) of the particle's worldlines. A
\vague{shock-wave-type} singularity
at the particle's creation light cone
can be understood similarly to a Cauchy horizon instability
inside a black hole
(see, e.g., Ref.~\cite{BurkoOri:book}); an observer crossing
the creation light cone sees an
infinitely long history of the source
in a finite interval of proper time.
In the scalar field case (not considered by
Penrose) no \vague{Gaussian-type} constraints exist and
retarded fields can be constructed. However, we
shall see that the strength of
the retarded field (the gradient of the field) of
a scalar monopole has a $\delta$-function character
on the creation light cone so that, for example,
its energy-momentum tensor cannot be evaluated there.
In the electromagnetic case
it is not even possible to construct
a purely retarded field of
a single monopole --- one has to allow
additional sources on the creation light cone to find
a consistent retarded solution vanishing outside the future
domain of influence of the sources.

In both our somewhat different explanation of the nonexistence
of purely retarded fields of general sources, and in
Penrose's original discussion, the main cause of difficulties is the
spacelike character of $\scry^-$
and the consequential existence of
the past horizons, respectively, \vague{creation light cones.}

It was only after we constructed the various types of
fields produced by sources on de~Sitter background
and analyzed their behavior when we
noticed Penrose's general
considerations in Ref.~\cite{Penrose:1964}. Our
original motivation has been to
understand fields of accelerated
sources, and in particular, the electromagnetic field of uniformly
accelerated charges in de~Sitter spacetime. The question of
electromagnetic field and its radiative properties produced by a
charge with hyperbolic motion in Minkowski spacetime has perhaps
been one of the most discussed
\vague{perpetual problems} of classical
electrodynamics if not of all classical physics in the 20th
century. Here let us only notice that the December~2000 issue of
Annals of Physics contains the series of three papers (covering
80 pp.) by Eriksen and Gr{\o}n
\cite{EriksenGron:2000}, which study in depth and
detail various aspects of ``electrodynamics of hyperbolically
accelerated charges''; the papers also contain many (though not
all) references on the subject.

The electromagnetic field of
a uniformly accelerated charge along the
$z$ axis, say, is symmetrical not only with respect to the
rotations around the axis, but also with respect to the boosts
along the axis. Now spacetimes with boost-rotation symmetry play
an important role in full general relativity (see, e.g.,
Ref.~\cite{Bicak:Ehlers}, and references therein).
They represent the only explicitly known exact
solutions of the Einstein vacuum field equations, which describe
moving \vague{objects} ---
accelerated singularities or black holes ---
emitting gravitational waves, and which are asymptotically flat
in the sense that they admit global, though not complete, smooth
null infinity $\scry^-$. Their radiative character is best
manifested in a nonvanishing
Bondi's news function, which is an
analog of the radiative part of the Poynting vector in
electrodynamics. The general structure of all vacuum
boost-rotation symmetric spacetimes with hypersurface orthogonal
Killing vectors was analyzed
in detail in Ref.~\cite{BicakSchmidt:1989}.
One of the best known examples
is the $C$-metric, describing uniformly
accelerated black holes attached to conical singularities
(\vague{cosmic strings} or
\vague{struts}) along the axis of symmetry.

There exists also the generalization of the $C$-metric including a
non\-van\-ish\-ing~$\Lambda$ \cite{PlebanskiDemianski:1976}.
It has been used to study the pair creation of black holes
\cite{MannRoss:1995}; its interpretation
as uniformly accelerating
black holes in a de~Sitter space has been discussed recently
\cite{PodolskyGriffiths:2001}. However, no general framework is
available to analyze the whole class of boost-rotation symmetric
spacetimes, which are asymptotically approaching a de~Sitter (or
anti--de~Sitter) spacetime as it
is given in Ref.~\cite{BicakSchmidt:1989} for
$\Lambda=0$. Before developing such a framework in full general
relativity, we wish to gain an understanding of fields produced
by (uniformly) accelerated sources in a de~Sitter background. This
has been our original motivation for this work.

Although it has been widely known and used in various contexts
that there exists a conformal transformation between de~Sitter
and Minkowski spacetimes, this fact does not seem to be employed
for constructing the fields
of specific sources. In the following
we make use of this conformal relation to find scalar and
electromagnetic fields of the scalar and electric charges in
de~Sitter spacetime.

The plan of the paper is as follows. In Section~\ref{sc:ConfInv},
we will analyze the behavior of scalar and electromagnetic field
equations with source terms under general conformal
transformations. Few points contained here appear to be new,
like the behavior of scalar sources in a general, $n$-dimensional
spacetime, but the main purpose of this section is to review
results and introduce notation needed in subsequent parts. In
Section~\ref{sc:MinkDSCompact}, the compactification
of Minkowski and de~Sitter
spacetimes and their conformal properties are discussed. Again,
all main ideas are known, especially from works of Penrose. But
we need the detailed picture of the complete compactification of
both spaces and explicit formulas connecting them in various
coordinate systems, in order to be able to \vague{translate}
appropriate motion of the sources
and their fields from Minkowski into
de~Sitter spacetime. The worldlines of uniformly accelerated
particles in de~Sitter space are defined, found, and their
relation to the corresponding
world\-lines in Minkowski space under
the conformal mapping is
discussed in Section~\ref{sc:PointPart}.
In general, a single worldline in
Minkowski space gets transformed
into two worldlines in de~Sitter space.

In Section~\ref{sc:SymField}, by using
the conformal transformation of simple
boosted spherically symmetric fields of sources in Minkowski
spacetime, we construct the fields of uniformly accelerated
monopole sources in de~Sitter spacetime.
In particular, with both the scalar
and electromagnetic fields, we
obtain what we call \vague{symmetric fields.} They are analytic
everywhere outside the sources and can be written as a linear
combination of retarded and advanced fields
from both particles. From the symmetric fields we wish
to construct purely retarded fields
that are nonvanishing only in
the future domain of influence of
particles' worldlines. For the
scalar field, this is accomplished
in Section~\ref{sc:SFRetField}. We do find the retarded field,
but its strength contains a
$\deltafc$-function term located on the
particle's past horizon (creation light cone). In
Section~\ref{sc:EMRetField}, the retarded
electromagnetic fields are analyzed
for free (unaccelerated) monopoles (Subsection~\ref{ssc:FreeMon}),
for \vague{rigid dipoles} (Subsection~\ref{ssc:RigDip}),
consisting of two close, uniformly accelerated charges of
opposite sign, and for \vague{geodesic dipoles}
(Subsection~\ref{ssc:GeodDip}), made of two free opposite charges moving
along geodesics. In Subsection~\ref{ssc:EMConstr}
the role of the constraints, which electromagnetic fields and
charges have to satisfy on any spacelike hypersurface, is
emphasized. These constraints in de~Sitter space with compact
spatial slicings require the total charge to be zero. As is
well known, there can be no net charge in a closed universe (see,
e.g., Ref.~\cite{MTW}). However, we find out that the constraints imply
even \emph{local} conditions on the charge distribution if
$\scry^-$ is spacelike and purely retarded fields are only
admitted. In the case of an unaccelerated electromagnetic
monopole, we discover that the solution resembling retarded field
represents not only the monopole charge but also a spherical
shell of charges moving with the velocity of light along the
creation light cone of the monopole. The total charge of the
shell is precisely opposite to that of the monopole. Retarded
fields of both rigid and geodesic dipoles blow up along the
creation light cone since, by restricting ourselves to the fields
nonvanishing in the future
domain of influence, we \vague{squeeze}
the field lines produced by the dipoles into their past horizon
(creation light cone).

We do not discuss the radiative character
of the fields obtained. The problem of radiation
is not a straightforward issue since the
conformal transformation does not map an infinity
onto an infinity and, thus, one has to analyze
carefully the falloff (\vague{the peeling off}) of the
fields along appropriate null geodesics going to
future, respectively, past spacelike infinity. A detailed
discussion of the radiative properties of
the solutions found here and of some additional
fields will be given in a forthcoming publication
\cite{BicakKrtous:BIS}.

A brief discussion in Section~\ref{sc:summary}
concludes the paper. Some details concerning coordinate
systems on de~Sitter space are relegated to the Appendix.

\section{Conformal invariance of scalar
and electromagnetic field equations with sources}
\label{sc:ConfInv}

Conformal rescaling of metric is given by a common spacetime
dependent conformal factor $\Omega(x)$:
\begin{equation}\label{CTmetric}
\mtrc_{\alpha\beta} \;\;\rightarrow\;\;
\hat\mtrc_{\alpha\beta}=
\Omega^2\mtrc_{\alpha\beta}\comma
\mtrc^{\alpha\beta} \;\;\rightarrow\;\;
\hat\mtrc^{\alpha\beta}=
\Omega^{\sminus 2}\mtrc^{\alpha\beta}\period
\end{equation}
An equation for a physical field $\Psi$ is called
conformally invariant if there exists a number
--- \defterm{conformal weight} ---
${p\in\realn}$ such that $\hat\Psi=\Omega^p\Psi$ solves
a field equation with metric $\hat\mtrc$, if
and only if $\Psi$ is a solution of the original equation
with metric~$\mtrc$.

It is well known (see, e.g., Ref.~\cite{Wald:book1984}) that (i) the
wave equation for a scalar field $\SF$ can be generalized in
a conformally invariant way to curved $n$-dimensional
spacetime geometry by a suitable coupling with
the scalar curvature $\sccr$, and (ii) the vacuum Maxwell's
equations are conformally invariant in four dimensions
with conformal weight ${p=0}$ of covariant components
(2-form) of electromagnetic field $\EMF_{\alpha\beta}$,
but they fail to be conformally invariant for dimensions $n\neq4$.

The behavior of the above
equations with sources is not so widely known
(cf. Refs.~\cite{FultonRohrlichWitten:1962,PenroseRindler:book} for
the electromagnetic case). It is, however, easy to see that the
wave equation for the scalar field $\SF$ with the scalar charge
source~$\SFS$,
\begin{equation}\label{SFWE}
[\dalamb-\SFx\sccr]\SF = \SFS\commae
\end{equation}
where in $n$ dimensions $\SFx=\frac{n-2}{4(n-1)}$,
$\sccr$ is the scalar curvature, and
$\dalamb=\mtrc^{\alpha\beta}\stcnx_\alpha\stcnx_\beta$
is the d'Alembertian constructed
from the covariant metric derivative~$\stcnx_\alpha$,
under the conformal rescaling Eq.~\eqref{CTmetric} goes
over into the equation of the same form
\begin{equation}
[\hat\dalamb-\SFx\hat\sccr]\hat\SF
= \hat\SFS\commae
\end{equation}
provided that
\begin{gather}
\SF \;\;\rightarrow\;\; \hat\SF=
\Omega^{1-\frac{n}{2}}\;\SF\commae\label{SFCF}\\
\SFS \;\;\rightarrow\;\;  \hat\SFS=
\Omega^{-1-\frac{n}{2}}\;\SFS\commae\label{SFSCT}
\end{gather}
and $\hat\stcnx_\alpha$ and $\hat\sccr$
are the metric covariant derivative and scalar
curvature associated with the rescaled metric $\hat\mtrc$
(see, e.g., Eqs. (D.1)--(D.14) in Ref.~\cite{Wald:book1984}).

Next, it is easy to demonstrate that in four dimensions
Maxwell's equations with a source
given by a four-current $\EMJ^{\alpha}$,
\begin{equation}
\begin{gathered}
\stcnx_\mu\EMF^{\alpha\mu} = \EMJ^\alpha\commae\\
\stcnx_{[\alpha} \EMF_{\beta\gamma]} = 0 \quad\text{or}\quad
\EMF_{\alpha\beta}=\stcnx_\alpha \EMA_\beta -
\stcnx_\beta \EMA_\alpha\commae
\end{gathered}
\end{equation}
are conformally invariant if the vector potential
does not change, so that
\begin{equation}\label{EMCT}
{\hat\EMA}_\alpha = \EMA_\alpha\comma
{\hat\EMF}_{\alpha\beta} = \EMF_{\alpha\beta}\comma
{\hat\EMF}^{\alpha\beta} =
\Omega^{\sminus 4}\,\EMF^{\alpha\beta}\commae
\end{equation}
and the current behaves as follows:
\begin{equation}\label{EMJCT}
{\hat\EMJ}^\alpha = \Omega^{\sminus 4}\,\EMJ^{\alpha}\comma
{\hat\EMJ}_\alpha = \Omega^{\sminus 2}\,\EMJ_{\alpha}\period
\end{equation}
Since the Levi-Civita tensor transforms as
\begin{equation}\label{LCtCT}
{\hat\LCt}_{\alpha\beta\gamma\delta} =
\Omega^4 \LCt_{\alpha\beta\gamma\delta}\commae
\end{equation}
the following quantities are conformally invariant:
\begin{gather}
\dual{\hat\EMJ}_{\alpha\beta\gamma} =
{\hat\LCt}_{\alpha\beta\gamma\mu} {\hat\EMJ}^{\mu}
= \dual\EMJ_{\alpha\beta\gamma}\commae\\
\dual{\hat\EMF}_{\alpha\beta} = \frac{1}{2!}
{\hat\LCt}_{\alpha\beta\mu\nu} {\hat\EMF}^{\mu\nu}
= \dual\EMF_{\alpha\beta}\period
\end{gather}
Therefore, Maxwell's equations with a source
can be written using the external derivative as
\begin{equation}
\grad\EMF = 0\comma
\grad\dual\EMF=-2\, \dual\EMJ\commae
\end{equation}
where only conformally invariant quantities appear.

The continuity equation for the electromagnetic current
is also conformally invariant:
\begin{equation}
\stcnx_\alpha\EMJ^\alpha = 0 \;\;\rightarrow\;\;
{\hat\stcnx}_\alpha{\hat\EMJ}^\alpha =
\Omega^{\sminus 4} \stcnx_\alpha\EMJ^\alpha = 0
\end{equation}
thanks to Eq.~\eqref{EMJCT} and the conformal property of
the four-dimensional volume element
$\vol = (-\Det\mtrc_{\alpha\beta})^{1/2}$:
\begin{equation}
{\hat\stmtrcdet}^{1/2} = \Omega^4 \vol\period
\end{equation}
It is interesting to notice, however, that as
a consequence of the invariance \eqref{EMCT}
of the electromagnetic potential under a
conformal rescaling, the Lorentz gauge condition
is not conformally invariant:
\begin{equation}
\stcnx_\alpha\EMA^\alpha = 0
\end{equation}
implies
\begin{equation}
{\hat\stcnx}_\alpha{\hat\EMA}^\alpha
- 2 {\hat\EMA}^\alpha\, \grad_\alpha\log\abs{\Omega}
=0 \period
\end{equation}

A remarkable property arises in four-dimensional
spacetimes: in both the scalar and electromagnetic case
the total charge distributed on a three-di\-men\-sional
spacelike hypersurface is conformally
(pseudo)invariant.\footnote{
  A quantity is a conformal pseudoinvariant if
  it is invariant under conformal transformation,
  except for a change of sign if
  the conformal factor is negative.
  See Eq. \eqref{SFsCT}.}
This follows form the conformal invariance of
spatial charge distributions.

Denoting $\normal_\alpha$ a future-oriented
unit 1-form normal to the hypersurface $\Sigma$,
we get
\begin{equation}\label{normalCT}
{\hat\normal}_\alpha = \abs{\Omega}\normal_\alpha\comma
{\hat\normal}^\alpha = \abs{\Omega}^{-1}\normal^\alpha\period
\end{equation}
The three-dimensional volume element is given by
$\svol=(\Det\smtrc_{\alpha\beta})^{1/2}$,
where three-metric $\smtrc_{\alpha\beta}$ is the restriction
of the four-metric $\mtrc_{\alpha\beta}$ to the
hypersurface $\Sigma$. Under the conformal transformation,
\begin{equation}\label{svolCT}
\svol \;\;\rightarrow\;\;
{\hat\smtrcdet}{}^{1/2}=\abs{\Omega}^3\svol\period
\end{equation}
A charge distribution is defined as a charge
density multiplied by this volume element.
Hence, the scalar charge distribution reads
\begin{equation}
\SFs = \SFS\,\svol\commae
\end{equation}
and
\begin{equation}\label{SFsCT}
\SFs \;\;\rightarrow\;\;
\hat\SFs = \sign(\Omega)\; \SFs\period
\end{equation}
We see that it is conformal invariant except for a
change of sign if the conformal factor $\Omega$ is negative.
As seen from Eq.~\eqref{SFSCT}, this fails to be true for $n\neq4$.
In the following we consider only the case $n=4$.

The electromagnetic charge distribution is given by
\begin{equation}
\EMr = \normal_\alpha\EMJ^\alpha\;\svol\period
\end{equation}
Again, regarding Eqs.~\eqref{EMJCT}, \eqref{normalCT},
and \eqref{svolCT}, we get
\begin{equation}\label{EMrCT}
\EMr \;\;\rightarrow\;\;
\hat\EMr = \EMr\period
\end{equation}
Thus, the electromagnetic charge is invariant even
under conformal transformation with a negative
conformal factor $\Omega$.

Similarly, we define the electric field with
the three-dimensional volume element included:
\begin{equation}\label{EMEdef}
\EME^\alpha = \normal_\mu \EMF^{\mu\alpha}\;\svol\commae
\end{equation}
which represents the momentum conjugated
to the potential $\EMA_\alpha$ (cf., e.g., Ref. \cite{MTW}).
With the definition \eqref{EMEdef}, $\EME^\alpha$ is
conformally invariant. Gauss's law
simply reads
\begin{equation}
\int_{D}\scnx_\alpha \EME^\alpha =
\int_{\partial{D}}\EME^\alpha \bndrestr_\alpha\commae
\end{equation}
where $\scnx$ is three-metric covariant derivative
and $D$ is a region in $\Sigma$.

\section{Minkowski and de~Sitter spacetimes:\newline
compactification and conformal relation}
\label{sc:MinkDSCompact}

The conformal structure of Minkowski and de~Sitter spacetimes
and their conformal relation to the regions of the Einstein
static universe is well known and has been much used
(see, e.g., Refs.~\cite{HawkingEllis:book,Wald:book1984} for basic
expositions). However, the complete compactified picture of
both spaces and their conformal structure do not appear to be
described in detail in the literature, although all main
ideas are contained in various writings by Penrose
(e.g. Refs.~\cite{Penrose:1964,PenroseRindler:book}).
Since we shall need some details in explicit form when
analyzing the character of the fields of sources in de~Sitter
spacetime and their relation to their counterparts in
Minkowski spacetime, we shall now discuss the compactification
and conformal properties of these spaces.

Recall first that flat Euclidean plane $E^2$ can be compactified
by adding a point at infinity so that the resulting space is a
two-sphere $S^2$ with a regular homogenous metric $\mtrc_{\sphr}$,
conformally related to the Euclidean metric:
\begin{equation}
\mtrc_\sphr = \alpha^2 \bigl(\grad\vartheta\formsq+
\sin^2\!\vartheta\,\grad\varphi\formsq\bigr) =
\Omega^2 \bigl(\grad r\formsq + r^2 \grad\varphi\formsq\bigr) =
\Omega^2 \mtrc_\Eucl\commae
\end{equation}
where $\alpha$ is a constant parameter with the dimension
of length, $r=\alpha\tan\frac\vartheta2$,
and $\Omega=1+\cos\vartheta$. Notice that $\mtrc_\Eucl$ is
not regular at $r=\infty$, where the conformal factor
$\Omega=0$. The group of conformal transformations of
$E^2$ acts on the compact manifold $S^2$.

\begin{figure}[t]
  \begin{center}
  \epsfig{file=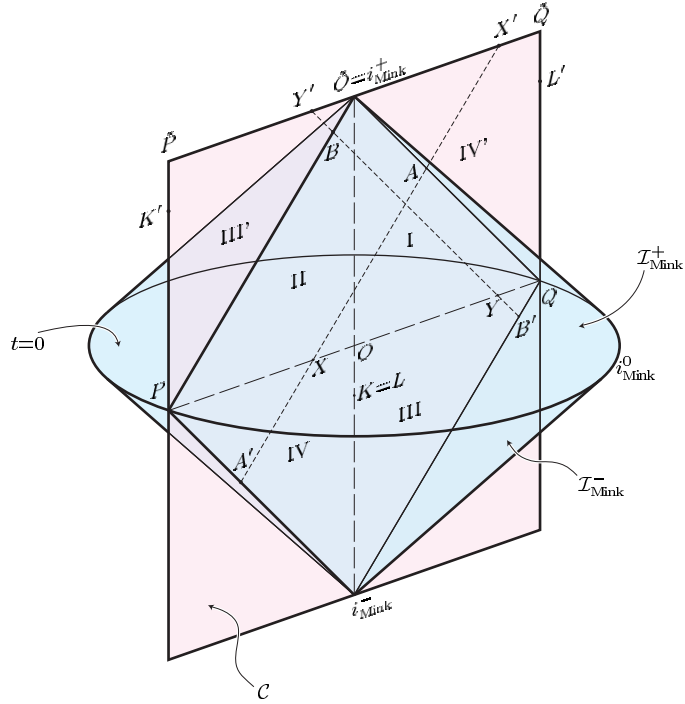}
  \end{center}
  \caption{Three-dimensional Penrose diagram of Minkowski space}
  \figdescription{
  The whole spacetime is mapped into the
  interior of two cones joined base to base along a
  spacelike (Cauchy) hypersurface $t=0$.
  The boundary of the two cones consists of past and
  future null infinities, $\scry_\Mink^-$, $\scry_\Mink^+$,
  of the past and future timelike infinities, $\stinf_\Mink^-$,
  $\stinf_\Mink^+$, and of the
  spacelike infinity, $\stinf_\Mink^0$.
  At these infinities, the null, timelike, and
  spacelike geodesics start and end.
  A two-dimensional cut $\mathcal{C}$
  going through $\stinf_\Mink^-$,
  $\stinf_\Mink^+$, and $P,Q\in\stinf_\Mink^0$
  is considered, with two null
  geodesics $A'XA$ and $B'YB$ indicated.
  It is divided into four separate regions, I--IV.
  Regions III and IV are mapped into regions
  III' and IV' in the compactified
  Minkowski space illustrated
  in Fig.~\ref{fig:dS}.
  }
  \label{fig:Mink}
\end{figure}

Analogously, one can construct compactified Minkowski space
$M^\#$ (see Ref. \cite{PenroseRindler:book}, Section~9.2 and
references therein) on which the 15-parameter
conformal group acts. One starts with the standard
Penrose diagram of Minkowski space
and makes an identification of
$\scry_\Mink^-$ and $\scry_\Mink^+$ by
identifying the past and future endpoints of null geodesics
as indicated in Fig.~\ref{fig:Mink}. The future and
past timelike infinities, $\stinf_\Mink^\pm$, and the spatial
infinity, $\stinf_\Mink^0$, are also identified into one point.
The topology of $M^\#$ is $S^3\times S^1$ (this is
not evident from first sight, but see
Ref.~\cite{PenroseRindler:book}).

\begin{figure}[t]
  \begin{center}
  \epsfig{file=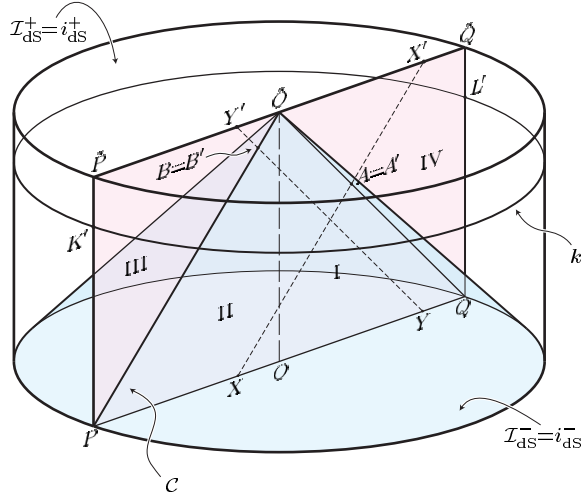}
  \end{center}
  \caption{Compactified Minkowski and de~Sitter spaces}
  \figdescription{The compactified
  Minkowski and de~Sitter space
  $M^\#$, illustrated by the three-dimensional diagram ---
  a part of the Einstein universe represented here
  as a solid cylinder. The compactification is achieved
  by considering first the two-dimensional section
  $\mathcal{C}$ in Fig.~\ref{fig:Mink},
  cutting out regions III and IV and placing them
  \vague{above} regions II and I so that
  two-dimensional figure $POQ\tilde{Q}\tilde{O}\tilde{P}$
  is formed. $PO$ is identified with $\tilde{O}\tilde{Q}$
  (e.g., point $X$ with $X'$), $OQ$ with $\tilde{P}\tilde{O}$
  (e.g., $Y$ with $Y'$), and $P\tilde{P}$ with $Q\tilde{Q}$.
  All two-dimensional cuts $\mathcal{C}$
  are identified in this way
  with, in addition, all \vague{vertical} boundary lines
  being identified so that the circle $k$, for example,
  is considered as a point. The top and bottom bases
  of the cylinder, representing the past and the future
  spacelike infinities of de~Sitter space, are identified in
  the compact manifold $M^\#$.
  }
  \label{fig:dS}
\end{figure}

Rescaled Minkowski space (without the identification)
can be drawn in a two-dimensional diagram as a part of
the Einstein static universe, which is visualized by
a cylindrical surface imbedded in $E^3$ --- see,
e.g., Refs.~\cite{HawkingEllis:book,Wald:book1984}. However, in
Fig.~\ref{fig:dS} we illustrate the compactified
Minkowski space $M^\#$ by a \emph{three-dimensional} diagram
as a part of the Einstein universe represented by a
\emph{solid} cylinder in $E^3$.
This is achieved in
the following way: In Fig.~\ref{fig:Mink} the Minkowski
spacetime (with one dimension suppressed) is illustrated
as a region bounded by two cones joined base to base.
Now we take a two-dimensional cut $\mathcal{C}$ and
we compactify it by dividing it into four regions, I--IV,
as indicated in Fig.~\ref{fig:Mink}.
We cut out regions III and IV
and place them \vague{above} regions II and I so that
they are joined along their corresponding
null boundaries (e.g., points $B$, $B'$ and $A$, $A'$
become identical). Now the segment
$PO$ has to be identified with $\tilde{O}\tilde{Q}$ and
$OQ$ with $\tilde{P}\tilde{O}$ ---
they correspond to a single
segment in Fig.~\ref{fig:Mink}. Similarly,
boundaries $P\tilde{P}$ and $Q\tilde{Q}$
are identified, and as a result
a compact manifold is formed. Consider then
all posible cuts $\mathcal{C}$, i.e., \vague{rotate}
$\mathcal{C}$ around the \vague{line}
$\stinf_\Mink^-O\stinf_\Mink^+$, and make the same
identifications as we just described. Now all
\vague{vertical} boundary lines
as $P\tilde{P}$ and $Q\tilde{Q}$
have to be identified (notice that all these points were
on the segment $\stinf_\Mink^-O$ in Fig.~\ref{fig:Mink}).
The resulting four-dimensional
compact manifold is represented in the three-dimensional
Fig.~\ref{fig:dS}. From the construction described,
it follows that the top and bottom bases of the solid
cylinder are identified and each
of the circles on the cylindrical
surface, as, e.g., $k$, should be considered
as a single point. The \vague{disks} inside these circles
are thus two-spheres, i.e., without suppressing  one dimension ---
three-spheres in $M^\#$.

Now it is important to realize that Fig.~\ref{fig:dS}
can be understood as a part of the Einstein static cylinder,
which also represents the \emph{compactified de~Sitter space}.
In de~Sitter space two bases are future and past spacelike
infinities. They are not usually identified in the standard
two-dimensional Penrose diagram of de~Sitter spacetime
(see, e.g., Ref.~\cite{HawkingEllis:book}),
as $\scry_\Mink^+$ and $\scry_\Mink^-$
are not identified in the standard Penrose
diagram of Minkowski space.

Manifold $M^\#$ represents
the compactification of both Minkowski
and de~Sitter space. Similarly as $S^2$, representing the
compactification of $E^2$, can be equipped with a regular
metric $\mtrc_\sphr$ mentioned above, $M^\#$ can be
equipped with the regular metric
\begin{equation}\label{EinstMtrc}
\mtrc_\Eins = \alpha^2\;\bigl(
-\grad\tlt\formsq+\grad\tlr\formsq+
\sin^2\!\tlr\;\sphmtrc\bigr)\commae
\end{equation}
where dimensionless coordinates
$\tlt,\tlr\in\langle0,\pi\rangle$,
spacelike hypersufaces $\tlt=0$, and $\tlt=\pi$ are identified
by means of null geodesics,
$\sphmtrc = \grad\vartheta\formsq +
\sin^2\!\vartheta\,\grad\varphi\formsq$,
and the constant $\alpha$ has dimension of length.
The metric \eqref{EinstMtrc} is the well known
metric of the Einstein universe, in which case
\begin{equation}\label{AlphaLambda}
\alpha^2=\frac{3}{\Lambda}\commae
\end{equation}
where $\Lambda$ is the cosmological constant.

In order to see this explicitly, write the Minkowski
metric in standard spherical coordinates,
\begin{equation}
\mtrc_\Mink =
-\grad t\formsq+\grad r\formsq + r^2\,\sphmtrc\commae
\end{equation}
and introduce coordinates $\tlt,\tlr\in\langle0,\pi\rangle$
by
\begin{equation}\label{tlttlrTtr}
\tlt = \arctan\frac{2\alpha t}{\alpha^2-t^2+r^2}\comma
\tlr = \arctan\frac{2\alpha r}{\alpha^2+t^2-r^2}\commae
\end{equation}
inversely
\begin{equation}\label{trTtlttlr}
t = \frac{\alpha \sin\tlt}{\cos\tlr+\cos\tlt}\comma
r = \frac{\alpha \sin\tlr}{\cos\tlr+\cos\tlt}\commae
\end{equation}
so that
\begin{equation}\label{MinkEinst}
\mtrc_\Mink =
\frac{\alpha^2}{(\cos\tlr+\cos\tlt)^2}
\;\bigl(
-\grad\tlt\formsq+\grad\tlr\formsq+
\sin^2\!\tlr\;\sphmtrc\bigr)\period
\end{equation}
Let us notice that by requiring the ranges
$\tlt,\tlr\in\langle0,\pi\rangle$,
we fix the branch of $\arctan$ in Eq.~\eqref{tlttlrTtr}.
Further, observe that for $\tlt+\tlr>\pi$,
relations \eqref{trTtlttlr}
imply negative $r$ --- we shall return to this point
in a moment.

In the case of de~Sitter space with the metric
\begin{equation}
\mtrc_\dS = -\grad\tau\formsq +
\alpha^2 \cosh^2\!\frac\tau\alpha\;
\Bigl(\grad\chi\formsq
+\sin^2\!\chi\;\sphmtrc\Bigr)\commae
\end{equation}
we put
\begin{equation}\label{tlttlrTtauchi}
\tlt = 2\arctan\Bigl(\exp\frac\tau\alpha\,\Bigr)\comma
\tlr = \chi\commae
\end{equation}
or
\begin{equation}\label{tauchiTtlttlr}
\tau = \alpha\log\Bigl(\tan\frac\tlt2\,\Bigr)\comma
\chi = \tlr\commae
\end{equation}
so that
\begin{equation}\label{dSEinst}
\mtrc_\dS =
\frac{\alpha^2}{\sin^2\tlt}
\;\bigl(
-\grad\tlt\formsq+\grad\tlr\formsq+
\sin^2\!\tlr\;\sphmtrc\bigr)\period
\end{equation}

In this way we obtain explicit forms of the conformal
rescaling of both spaces into the metric
of the Einstein universe:
\begin{align}
\mtrc_\Eins &= \Omega_\Mink^2\;\mtrc_\Mink\comma&
&\Omega_\Mink = \cos\tlr+\cos\tlt
\commae\label{MinkCF}\\
\mtrc_\Eins &= \Omega_\dS^2\;\mtrc_\dS\comma&
&\Omega_\dS = \sin\tlt
\commae\label{dSCF}
\end{align}
where $\mtrc_\Eins$ is given by Eq.~\eqref{EinstMtrc}.
As in the simple case of conformal relation
of $E^2$ to $S^2$, the conformal factors $\Omega_\Mink$
and $\Omega_\dS$ vanish at infinities of Minkowski,
respectively, de~Sitter space.

As a consequence of Eqs.~\eqref{tlttlrTtr}--\eqref{dSEinst}
we also find the conformal relation between Minkowski
and de~Sitter space:
\begin{equation}
\mtrc_\Mink = \Omega^2\mtrc_\dS\comma
\Omega=\Omega_\Mink^{\sminus1}\,\Omega_\dS\commae
\end{equation}
where $\Omega_\Mink$, $\Omega_\dS$ are given by
Eqs.~\eqref{MinkCF} and \eqref{dSCF}.
The conformal factor $\Omega$ has the
simplest form when expressed in terms of the
Minkowski time $t$:
\begin{equation}\label{CFfact}
\Omega=\frac{\sin\tlt}{\cos\tlr+\cos\tlt}
= \frac{t}{\alpha}\period
\end{equation}
The conformal transformation is not regular
at the infinity of Minkowski, respectively,
de~Sitter space because $\Omega$ diverges,
respectively, vanishes there. We shall return to
this point at the end of this section. First, however,
we have to describe the coordinate systems employed in
relating particular regions I--IV in Figs.~\ref{fig:Mink}
and \ref{fig:dS}.

Relations \eqref{tlttlrTtr} and \eqref{trTtlttlr}
can be used automatically in region I only.
In other regions, ranges of coordinates have to
be specified more carefully.
In the following we always
require $\tlt\in\langle0,\pi\rangle$.
Then, if $\tlr\in\langle0,\pi\rangle$, we find that
relations \eqref{tlttlrTtr} and \eqref{trTtlttlr} imply
negative $r$ in region IV. Also, if we consider
events with $t<0$, $r>0$ (region III), we notice
that as a consequence of
Eqs.~\eqref{tlttlrTtr} and \eqref{trTtlttlr} with
$\tlt\in\langle0,\pi\rangle$ we get $\tlr<0$.
Relations \eqref{tlttlrTtr} and \eqref{trTtlttlr}
can be made meaningful in all
regions I--IV if we allow negative $r$, $\tlr$ and
adopt the following convention:
at a fixed value of time coordinate $t$, respectively, $\tlt$,
the points symmetrical with respect
to the origin of spherical coordinates
have opposite signs of the radial coordinate,
i.e., points with given
$\{t,r,\vartheta,\varphi\}$, respectively
$\{\tlt,\tlr,\vartheta,\varphi\}$, are
identical with
$\{t,-r,\pi-\vartheta,\varphi+\pi\}$, respectively
$\{\tlt,-\tlr,\pi-\vartheta,\varphi+\pi\}$.
The way in which regions I--IV are covered
by the particular ranges of coordinates
is explicitly illustrated in Figs.~\ref{fig:Coord}(a)--\ref{fig:Coord}(c)
in the Appendix, where our convention is described in more detail.

In the Appendix, various useful coordinate
systems in de~Sitter space are given. First,
we shall frequently employ coordinates
$\{\tlt,\tlr,\vartheta,\varphi\}$ which are
simply related (by Eqs.~\eqref{tlttlrTtauchi}
and \eqref{tauchiTtlttlr}) to the standard
coordinates $\{\tau$, $\chi$, $\vartheta$, $\varphi\}$
covering nicely the whole de~Sitter
hyperboloid. Next, relations
\eqref{tlttlrTtr} and \eqref{trTtlttlr}
can be viewed as the definition of
another coordinate system $\{t,r,\vartheta,\varphi\}$
on de~Sitter space (with the metric being given by
Eq.~\eqref{dScoorConf}).
Let us remind that for fixed $\vartheta$, $\varphi$
values $\tlr>0$ (commonly assumed in
de~Sitter space) correspond to $r>0$ for $\tlt+\tlr<\pi$
(region I) and to $r<0$ for $\tlt+\tlr>\pi$
(region IV). Further, one frequently uses
static coordinates $T$, $R$ associated with
the static Killing vector of de~Sitter spacetime ---
Eqs.~\eqref{dScoorStat} in the Appendix.

Finally, it will be useful to introduce
the null coordinates (cf. Eq.~\eqref{dScoorNull})
\begin{gather}
u = t-r \comma v = t+r\commae\label{dScooruv}\\
\tlu = \tlt-\tlr \comma \tlv = \tlt+\tlr
\period\label{dScoortluv}
\end{gather}
When employing null coordinates we shall consider
only $\tlr>0$ ($\tlu$, $\tlv$ would \vague{exchange
their role} if $\tlr<0$); the ranges of
$\tlu,\tlv$ and $u,v$ are thus given by the choice
$\tlt,\tlr\in\langle0,\pi\rangle$.
This leaves the standard (Minkowski) meaning of
$u$, $v$ in region I; however, $u$ and $v$ exchange
their usual (Minkowski) role in region IV
($\tlv=\tlt+\tlr>\pi$) because here $r<0$.
With this choice, the coordinates $\tlu$, $\tlv$
and $u$ cover de~Sitter space continuously,
in particular the horizon $\tlv=\pi$
($v\rightarrow\pm\infty$ on this horizon
--- see Fig.~\ref{fig:Coord}(e)).

From Eqs.~\eqref{tlttlrTtr} and \eqref{trTtlttlr}
we get simple relations
\begin{equation}\label{uvTtlutlv}
u = \alpha \tan\frac\tlu2 \comma
v = \alpha \tan\frac\tlv2 \commae
\end{equation}
which explicitly verify that local
null cones (local causal structure) are
unchanged under conformal mapping.
Nevertheless, it is well known that the global
causal structure of the Minkowski and de~Sitter
space is different. This is reflected in the fact
that, as mentioned above, the conformal
transformation between the two spaces
is not regular everywhere. In particular,
relation \eqref{uvTtlutlv} shows that points
at null infinity with $v\rightarrow\pm\infty$
in Minkowski space go over
into regular points with $\tlv\rightarrow\pi$ in
de~Sitter space, whereas spacelike hypersurface
$t=\frac12(u+v)=0$ goes into spacelike
infinities $\tlt=0,\pi$ in de~Sitter space.

In the next sections, when we shall generate
solutions for the scalar and electromagnetic
fields for given sources in de~Sitter
space by employing the conformal transformation
from Minkowski space, we have to check
the behavior of the new solutions at points where
the transformation is not regular.

Before turning to the construction of the fields
produced by specific sources, let us emphasize
that in all the following expressions for
fields in de~Sitter spacetime only positive $\tlr$
can be considered. However, the results
contained in Sections~\ref{sc:PointPart}
and \ref{sc:SymField} are valid also for $\tlr<0$
provided that the convention described above is used.\footnote{
  In Section~\ref{sc:EMRetField}, we require $\tlr>0$.
  The right-hand sides of expressions
  \eqref{EMAsym-mon}, \eqref{EMFsym-mon},
  \eqref{EMsym-rdip}, and \eqref{EMsym-gdip} would
  have to be multiplied by a factor
  $\sign\tlr$ to be also valid for $\tlr<0$.
  Similar changes would also be necessary in
  other equations but these contain null
  coordinates that have not been defined for $\tlr<0$.}

\section{Uniformly accelerated particles
in de~Sitter spacetime}
\label{sc:PointPart}

In this section we study the correspondence of
the worldlines of uniformly accelerated particles under
the conformal mapping \eqref{tlttlrTtr} and \eqref{trTtlttlr}
between Minkowski and de~Sitter spacetimes.
Let a particle have four-velocity $u^\alpha$,
${u^\mu u_\mu = -1}$, so that its acceleration
is $a^\alpha = \dot{u}^\alpha = u^\mu \stcnx_\mu u^\alpha$,
$a^\mu u_\mu = 0$. We say that the particle
is uniformly accelerated if the projection
of $\dot{a}^\alpha = u^\mu \stcnx_\mu a^\alpha$
into the three-surface orthogonal to $u^\alpha$
vanishes:
\begin{equation}\label{UnifAcc}
P^\alpha_\mu\, \dot{a}^\mu =
\dot{a}^\alpha - (a^\mu a_\mu)\, u^\alpha = 0\period
\end{equation}
Here the projection tensor
$P^\mu_\nu = \delta^\mu_\nu + u^\mu u_\nu$
and $u_\mu \dot{a}^\mu = - a_\mu a^\mu$.
Multiplying Eq.~\eqref{UnifAcc} by $a^\alpha$,
we get $\dot{a}^\mu a_\mu = 0$ so that
\begin{equation}
a_\mu a^\mu = \text{constant}\period
\end{equation}
This definition of uniform acceleration goes
over into the standard definition used in Minkowski
space \cite{Rohrlich:book}. It implies
that the components of a particle's
acceleration in its instantaneous
rest frames remain constant.
Of course, as a special case, a particle
may have zero acceleration when it
moves along the geodesic.

\begin{figure}[t]
  \begin{center}
  \epsfig{file=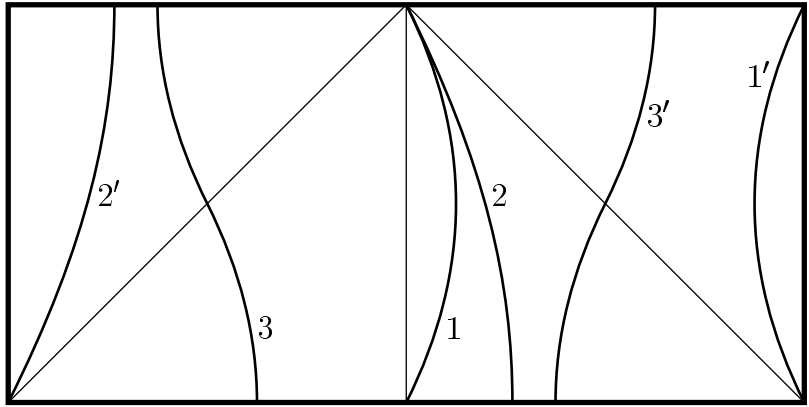}
  \end{center}
  \caption{Worldlines of particles in de~Sitter spacetime}
  \figdescription{The worldlines of geodesics and
  of uniformly accelerated particles in de~Sitter
  spacetime, obtained by the conformal transformation
  of appropriate worldlines in Minkowski space:
  $1$, $1'$ from the worldline of a particle
  moving uniformly through the origin,
  $2$, $2'$ from a particle at rest outside
  the origin, and $3$, $3'$ from two uniformly accelerated
  particles. In de~Sitter space, the worldlines $1$
  and $1'$ describe two uniformly accelerated
  particles; $2$, $2'$ and $3$, $3'$
  are geodesics. Both particles in each pair
  are causally disconnected.
  }
  \label{fig:Worldlines}
\end{figure}

Consider a particle moving with a constant
velocity
\begin{equation}
\frac{R_\oix}{\alpha} = \tanh\beta = \text{constant}
\end{equation}
along the $z$ axis ($\varphi=0$, $\vartheta=0$)
of the inertial frame in Minkowski spacetime
with coordinates $\{t,r,\vartheta,\varphi\}$
so that it passes through $r=0$ at $t=0$.
Transformations \eqref{tlttlrTtr}, \eqref{trTtlttlr}
map its worldline into \emph{two} worldlines
in de~Sitter spacetime, given in terms of
parameter $\lambda_\Mink$, its proper time
in Minkowski space, or in terms of $\lambda_\dS$,
its proper time in de~Sitter space, as follows:
\begin{equation}\label{AccMon}
\begin{gathered}
\tlt = \arctan\Bigl(-2\alpha\,
\frac{\lambda_\Mink\cosh\beta}{\lambda_\Mink^2-\alpha^2}
\Bigr) =
\arctan\Bigl(-\frac{\cosh\beta}{\sinh\bigl(
({\lambda_\dS}/{\alpha})\cosh\beta\bigr)}
\Bigr)\commae\\
\tlr = \arctan\Bigl(2\alpha\,
\frac{\lambda_\Mink\sinh\beta}{\lambda_\Mink^2+\alpha^2}
\Bigr) =
\arctan\Bigl(\pm\frac{\sinh\beta}{\cosh\bigl(
({\lambda_\dS}/{\alpha})\cosh\beta\bigr)}
\Bigr)\period
\end{gathered}
\end{equation}
In these expressions, the $\arctan$ has values in\footnote{
  $\tlr\in\langle-\pi,0\rangle$ for $\beta<0$.}
$\langle 0,\pi\rangle$;
$\lambda_\dS\in\realn$, and
$\lambda_\Mink\in(0,\infty)$ for the worldline
starting and ending with $\tlr=0$ (denoted
by $1$ in Fig.~\ref{fig:Worldlines}; plus
sign in the last equation in \eqref{AccMon}),
whereas $\lambda_\Mink\in(-\infty,0)$ for
the second worldline, starting and ending with
$\tlr=\pi$ (denoted by $1'$ in
Fig.~\ref{fig:Worldlines}; minus sign in
\eqref{AccMon}). One thus gets two worldlines
in de~Sitter space from two \vague{halves}
of one worldline in Minkowski space.

These two worldlines are uniformly accelerated
with the constant magnitude of the acceleration
equal (up to the sign) to
\begin{equation}
a_\oix = -\frac1\alpha\sinh\beta\period
\end{equation}
An intuitive understanding of the acceleration
is gained if we introduce standard \emph{static}
coordinates $\{T,R,\vartheta,\varphi\}$ in
de~Sitter space (see the Appendix).
The two worldlines described by Eq.~\eqref{AccMon}
in coordinates $\tlt,\tlr$ are in the static coordinates
simply given by $R=R_\oix=\text{constant}$.
(As seen from Eq.~\eqref{dScoorStat}, for a given
$\tlt$, $\tlr$, the same $R$ corresponds to
$\tlr$ and $\pi-\tlr$.) Owing to the
\vague{cosmic repulsion} caused by the presence of
a cosmological constant, fundamental geodesic
observers with fixed $\tlr$ (i.e., fixed $\chi$) are
\vague{repelled} one from the other in
proportion to their distance. Their initial
implosion starting at $\tlt=0$ ($\tau=-\infty$)
is stopped at $\tlt=\pi/2$ ($\tau=0$ ---
at the \vague{neck} of the de~Sitter hyperboloid)
and changes into expansion. A particle having
constant $R=R_\oix$, thus a constant proper distance
from an observer at $\tlr=R=0$ (or
at $\tlr=\pi$, $R=0$), has to be accelerated towards
that observer. The acceleration of particle $1$ points
towards the observer at $\tlr=0$, whereas that of
particle $1'$ points towards the observer at $\tlr=\pi$
(Fig.~\ref{fig:Worldlines}). Notice that the
two uniformly accelerated worldlines are causally
disconnected; no retarded or advanced effects from the
particle $1$ can reach the particle $1'$ and vice versa.
This is analogous to two particles symmetrically
located along opposite parts of say the $z$ axis
and uniformly accelerated in opposite directions
in Minkowski space. The worldlines of uniformly
accelerated particles in Minkowski space
are the orbits of the boost Killing vector.
Analogously, in de~Sitter space the Killing vector
$\cvect{T}$ also has the character of a boost.

Another type of simple worldline in de~Sitter
space arises from transforming the worldlines
of a particle at rest at $r=r_\oix$ ($\abs{r_\oix}<\alpha$),
$\vartheta=0$, $\varphi=0$ in Minkowski space.
It transforms to two worldlines in de~Sitter space given by
\begin{equation}\label{RigMon}
\begin{gathered}
\tlt = \arctan\Bigl(\frac{2\alpha\lambda_\Mink}
{-\lambda_\Mink^2+ r_\oix^2+\alpha^2}\Bigr)\commae\\
\tlr = \arctan\Bigl(\frac{-2\alpha r_\oix}
{-\lambda_\Mink^2+r_\oix^2-\alpha^2}\Bigr)\commae\\
\vartheta=0\comma\varphi=0\commae
\end{gathered}
\end{equation}
with $\tlt\in\langle0,\pi\rangle$, and
$\tlr\,\sign r_\oix\in\langle0,\pi\rangle$ for $\lambda_\Mink>0$
and $\tlr\,\sign r_\oix\in\langle-\pi,0\rangle$
for $\lambda_\Mink<0$.
Thus, a geodesic in Minkowski space goes over
into two geodesics in de~Sitter space.
In Fig.~\ref{fig:Worldlines} these are the worldlines
$2$ and $2'$.

As the last example, let us just mention that
the worldlines of two particles uniformly
accelerated in Minkowski space get transformed
into two geodesics $3$, $3'$ in de~Sitter space.

\section{Scalar and electromagnetic fields from
uniformly accelerated particles: the symmetric
solutions}
\label{sc:SymField}

Two uniformly accelerated particles described by
worldlines \eqref{AccMon} were obtained by the
conformal transformation from the worldline of a particle
moving with a uniform velocity ${R_\oix}/{\alpha}$
in Minkowski space. Hence, their fields can be constructed by
the conformal transformation of a simple boosted
spherically symmetric field. In the case of scalar
field, Eqs.~\eqref{SFCF}, \eqref{SFsCT}, and \eqref{CFfact}
then lead to the field
\begin{equation}\label{SFcol}
\SF = \frac{\SFq}{4\pi}\frac{t}{\alpha}
\,\frac{1}{\abs{r'}}\commae
\end{equation}
where $r'$ is spherical coordinate in an inertial
frame in which the particle is at rest at the origin.

As emphasized earlier, we have to examine the field
at the null hypersurface $\tlv=\pi$, where the conformal
transformation fails to be regular. We find that
field~\eqref{SFcol} is indeed not smooth there. The limit
of $\SF$ as $\tlv=\pi$ is approached from the region
$\tlv<\pi$ differs from the limit from $\tlv>\pi$;
$\SF$ has a jump at $\tlv=\pi$, although,
as can be checked by a direct calculation,
this discontinuous field satisfies the scalar
wave equation. However, a field analytic everywhere
outside the sources can be obtained by an analytic
continuation of the field \eqref{SFcol} from
the domain $\tlv<\pi$ to the domain $\tlv>\pi$.
We discover that the new field in $\tlv>\pi$
differs from Eq.~\eqref{SFcol} just by a sign. Therefore
it simply corresponds to the charge of the particle
in $\tlv>\pi$, which is opposite to that implied by
conformal transformation. It is easy to see
that, due to the conformal transformation,
the sign of the charge on the worldline with $\tlv>\pi$ is
opposite to the original charge $\SFq$ because
the conformal factor $\Omega=t/\alpha<0$
for $\tlv>\pi$. Hence, the field which is analytic
represents the field of two uniformly accelerated
particles with the \emph{same} scalar charge $\SFq$,
which move along two worldlines given by \eqref{AccMon}.
In Section~\ref{sc:SFRetField} we shall see
that this field can be written as a linear combination
of retarded and advanced fields from both
particles. We call it the \defterm{symmetric field}.
Regarding Eq.~\eqref{SFcol}, in which $r'$ is first expressed
in terms of the original Minkowski coordinates
$\{t,r,\vartheta,\varphi\}$, and then using the transformation
\eqref{trTtlttlr}, we find the field as a function
of $\{\tlt,\tlr,\vartheta,\varphi\}$:
\begin{equation}\label{SFsym-mon}
\SF_\sym = \frac{\SFq}{4\pi}
\frac{\sqrt{\alpha^2-R_\oix^2}}
{\sqrt{(\alpha^2-R R_\oix\cos\vartheta)^2-
(\alpha^2-R^2)(\alpha^2-R_\oix^2)}}\commae
\end{equation}
where $R=\alpha\frac{\sin\tlr}{\sin\tlt}$
is the static radial de~Sitter coordinate
(see the Appendix). As could
have been anticipated, the field is static in the
static coordinates since the accelerated
particles are at rest at $R=R_\oix$, $\vartheta=\varphi=0$.
(Recall that we need two sets of such coordinates
to cover both worldlines but the coordinate $R$
is well defined in the whole de~Sitter spacetime
--- cf. the Appendix.) However,
it is dynamical in the coordinates
$\{\tlt,\tlr,\vartheta,\varphi\}$,
or in the standard coordinates
$\{\tau,\chi,\vartheta,\varphi\}$,
covering --- in contrast to the static coordinates ---
the whole de~Sitter spacetime.

In order to construct the electromagnetic field
produced by uniformly accelerated particles in
de~Sitter spacetime, we start, analogously to the
scalar field case, from the boosted Coulomb
field in Minkowski space. The potential 1-form
is thus simply
\begin{equation}\label{EMcol}
\EMA = -\frac{\EMq}{4\pi}\frac{1}{\abs{r'}}
\,\grad t'\commae
\end{equation}
where the prime again denotes the coordinates
in an inertial frame in which the particle
is at rest. Since electromagnetic
field described by its covariant component is
conformally invariant, the field \eqref{EMcol}
\emph{is} automatically a solution of Maxwell's
equations in de~Sitter space. However, like
in the scalar field case, we have to examine its character
at $\tlv=\pi$. We discover that the potential \eqref{EMcol}
does \emph{not}, in fact, solve Maxwell's equations
there (in contrast to the scalar field \eqref{SFcol},
which is discontinuous on $\tlv=\pi$, but satisfies
the scalar wave equation). This result can be
understood when we realize that by the conformal
transformation the sign of the electric charge ---
in contrast to the scalar charge --- does \emph{not}
change at the worldline with $\tlv>\pi$ so that the total
electric charge is $2\EMq$. A nonzero total charge in
de~Sitter spacetime, however, violates the constraint,
as we shall see in Subsection~\ref{sc:EMRetField}.\ref{ssc:EMConstr}.
In fact, it is well known that in a closed
universe the total electric charge must be zero due to
the Gauss's law (e.g. Ref.~\cite{MTW}).

As with the scalar field, we still can construct
a field smooth everywhere outside the sources
by analytic continuation of the field
obtained in the region $\tlv<\pi$ across
$\tlv=\pi$ into whole spacetime. Similar to the
scalar case, the resulting field in $\tlv>\pi$
corresponds just to the opposite charge, so that now,
in the electromagnetic case, the total charge
is indeed zero. The electromagnetic field
can be written as a combination of retarded
and advanced fields from both charges, as will
be shown in Subsection~\ref{sc:EMRetField}.\ref{ssc:FreeMon}.
The potential describing this \defterm{symmetric field}
has a simple form in the static coordinates:
\begin{equation}\label{EMAsym-amon}
\begin{split}
\EMA_\sym = - \frac{\EMq}{4\pi\monfact}\Bigl(&
\bigl(1-\frac{R_\oix R}{\alpha^2}\cos\vartheta\bigr)\,
\alpha\,\grad T \\
& + \frac{R-R_\oix\cos\vartheta}{1-R^2/\alpha^2}\,\grad R +
R R_\oix \sin\vartheta \, \grad\vartheta\Bigr)\commae
\end{split}
\end{equation}
where\footnote{
  The positive root $\monfact>0$ is taken here
  --- as in Eq.~\eqref{SFsym-mon}.}
\begin{equation}\label{RetR}
\monfact^2 = (\alpha^2-R R_\oix \cos\vartheta)^2
-(\alpha^2-R_\oix^2)(\alpha^2-R^2)\period
\end{equation}
As noticed in Section~\ref{sc:ConfInv},
the Lorentz gauge condition is not conformally
invariant, so that the potential \eqref{EMAsym-amon}
need not satisfy the condition,
although the original Coulomb
field does. Expressing the static radial coordinate as
$R=\alpha\frac{\sin\tlr}{\sin\tlt}$, and similarly $T$
(see the Appendix), we can find
the potential in global coordinates
$\{\tlt,\tlr,\vartheta,\varphi\}$, respectively,
$\{\tau,\chi,\vartheta,\varphi\}$. Since the
resulting form is not simple, we do not write
it here, but we give the electromagnetic
field explicitly in both the static and global
coordinates. In static coordinates it reads
\begin{equation}\label{EMFsymRT-amon}
\begin{split}
\EMF = \grad\EMA &= -\frac{\EMq}{4\pi}\,
\frac{\alpha(\alpha^2-R_\oix^2)}{\monfact^3}\\\times
&\biggl((R-R_\oix\cos\vartheta)\; \grad T \wedge \grad R
+ \Bigl(1-\frac{R^2}{\alpha^2}\Bigr) R R_\oix \sin\vartheta \;
\grad T \wedge \grad\vartheta\biggr)\commae
\end{split}
\end{equation}
where $\monfact$ is given by Eq.~\eqref{RetR}.
In $\{\tlt,\tlr,\vartheta,\varphi\}$ coordinates
we explicitly find
\begin{equation}
\label{EMFsymtl-amon}\raisetag{29pt}
\begin{split}
\EMF = &-\frac{\EMq}{4\pi}\,
\frac{\alpha^2 - R_\oix^2}{\monfact^3}
\frac{\alpha^3}{\sin^3\tlt}
\;\Bigl(
(\alpha\sin\tlr-R_\oix \sin\tlt\cos\vartheta)\,
\grad\tlt\wedge\grad\tlr\\
& + R_\oix \sin\tlt\cos\tlr\sin\tlr\sin\vartheta\,
\grad\tlt\wedge\grad\vartheta
- R_\oix \cos\tlt\sin\tlr\sin\tlr\sin\vartheta\,
\grad\tlr\wedge\grad\vartheta\Bigr)\period
\end{split}
\end{equation}

Summarizing, the field \eqref{EMFsymtl-amon}
represents the time-dependent electromagnetic
field of two particles with charges $\pm\EMq$,
uniformly accelerated along the worldlines
\eqref{AccMon} with accelerations
$\mp\alpha^{-1}\sinh\beta=
\mp R_\oix/(\alpha\sqrt{\alpha^2-R_\oix^2}\,)$.
The field is analytic everywhere outside the
charges. In the static coordinates the charges are
at rest at $R=R_\oix$ and their static field
is given by Eq.~\eqref{EMFsymRT-amon}.

\section{Scalar field: the retarded solutions}
\label{sc:SFRetField}

The symmetric scalar field solution \eqref{SFsym-mon},
representing two uniformly accelerated scalar
charges, is non-vanishing in the whole de~Sitter
spacetime. As mentioned before, and will be proved
at the end of this section (see Eqs.~\eqref{symRAfield} and
\eqref{symHalfRAfield}), this field is a combination of
retarded and advanced effects from both charges.
A retarded field of a point particle should in general
be nonzero only in the future domain of influence of
a particle's worldline, i.e., at those points from
which past causal curves exist which intersect the
worldline. Hence, the retarded field of the uniformly
accelerated charge, which starts and ends at $\tlr=0$
(see Fig.~\ref{fig:Worldlines}), should be nonvanishing
only at $\tlu=\tlt-\tlr>0$. It is natural to try
to construct such a field by restricting the
symmetric field to this region, i.e., to ask
whether the field
\begin{equation}\label{SFret-amon}
\SF_\ret = \SF_\sym\,\stepfc(\tlu)\commae
\end{equation}
where $\stepfc$ is the usual Heaviside step function,
is a solution of the field equation.

The field \eqref{SFret-amon} does, of course, satisfy
the scalar field wave equation Eq.~\eqref{SFWE} at
$\tlu>0$ since $\SF_\sym$ does, and also at $\tlu<0$
since $\SF=0$ is a solution of \eqref{SFWE} outside
a source. Thus we have to examine the field
\eqref{SFret-amon} only at $\tlu=0$, i.e., at
\vague{creation light cone} of the particle's
worldline, also referred to as the past event horizon
of the worldline \cite{HawkingEllis:book}.
The field strength 1-form implied by Eq.~\eqref{SFret-amon}
becomes
\begin{equation}\label{dSF-amon}
\grad\SF_\ret = (\grad\SF_\sym)\,\stepfc(\tlu) +
\SF_\sym\;\deltafc(\tlu)\;\grad\tlu\period
\end{equation}
An explicit calculation shows that
\begin{equation}\label{SFretWE}
\dalamb\SF_\ret =
\bigl(\dalamb\SF_\sym\bigr)\stepfc(\tlu)\period
\end{equation}
Therefore, the conformally invariant scalar wave
equation \eqref{SFWE} (with $\SFx=\frac16$) has
the form
\begin{equation}
\Bigl[\dalamb-\frac16\sccr\Bigr]\SF_\ret =
\Bigl(\Bigl[\dalamb-\frac16\sccr\Bigr]\SF_\sym\Bigr)
\,\stepfc(\tlu) = \SFS_\sym\,\stepfc(\tlu)
=\SFS_{\mon1}\commae
\end{equation}
where $\SFS_{\mon1}$ denotes the monopole scalar
charge starting and ending at $\tlr=0$. Hence,
we proved that the field \eqref{SFret-amon},
where $\SF_\sym$ is given by Eq.~\eqref{SFsym-mon},
satisfies the field Eq.~\eqref{SFretWE}
everywhere, including the past event horizon of
the particle.

Analogously, we can make sure that
\begin{equation}\label{SFadv-amon}
\SF_\adv = \SF_\sym\,\stepfc(-\tlu)
\end{equation}
has its support in the future domain of influence
of the monopole particle $1'$, starting and ending
at $\tlr=\pi$, and is thus the advanced field of
source $\SFS_{\mon1'}=\SFS_\sym\,\stepfc(-\tlu)$.

From the results above it is not difficult to
conclude that the symmetric field can be
interpreted as arising from the combinations
of retarded and advanced potentials due to both
particles $1$ and $1'$, in which the potentials due
to one particle can be taken with arbitrary weights,
and the weights due the other particle then
determined by
\begin{equation}\label{symRAfield}
\SF_\sym = \zeta \SF_{\ret1} + (1-\zeta)\SF_{\adv1}
+(1-\zeta)\SF_{\ret1'}+\zeta\SF_{\adv1'}\commae
\end{equation}
where $\zeta\in\realn$ is an arbitrary constant
factor. In particular, choosing $\zeta=\frac12$,
the field
\begin{equation}\label{symHalfRAfield}
\SF_\sym = \frac12\bigl( \SF_\ret + \SF_\adv\bigr)
\end{equation}
is the symmetric field from both particles.
This freedom in the
interpretation is exactly the same as with
two uniformly accelerated scalar particles in
Minkowski spacetime (see Ref.~\cite{BicakSchmidt:1989},
Section~IV~B).

A remarkable property of the retarded field
\eqref{SFret-amon} is that the field strength
\eqref{dSF-amon} has a term proportional to
$\deltafc(\tlu)$, i.e., it is singular at
the past horizon. Since the energy-momentum
tensor of the scalar field is quadratic in the field
strength, it cannot be evaluated at $\tlu=0$.
The \vague{shock wave} at the \vague{creation
light cone} can be understood on physical grounds
similarly as the instability of Cauchy horizons
inside black holes (e.g., Ref.~\cite{BurkoOri:book});
an observer crossing the pulse along a timelike
worldline will see an infinitely long
history of the source within a finite proper time.
The character of the shock is given by the
pointlike nature of the source.
If, for example, a scalar charge has typical extension
$l$ at $\tlt=\pi/2$, i.e., at the moment of the
minimal size of the de~Sitter universe ($\tau=0$),
and the extension of the charge in $\tlr$ coordinate
would be roughly the same at the $\scry_\dS^-$, the
corresponding shock would be smoothed around $\tlu=0$
with a width $\sim l$. However, the \emph{proper} extension of
the charge at $\scry_\dS^-$ would be infinite in that case.

Let us note that the retarded field \eqref{SFret-amon}
could also be computed by means of the retarded
Green's function. In our case of the conformally
invariant equation for a scalar field, the retarded
Green's function in de~Sitter space
is localized on the future null cone,
as it is in the \vague{original} Minkowski space.
(It is interesting to note that in the case of a minimal,
or more general coupling, the scalar field
does not vanish inside the null cone.)
Thanks to this property we can understand a \vague{jump}
in the field on the creation light cone:
the creation light cone is precisely the future light
cone of the point at which the source \vague{enters}
the spacetime, i.e., it is the boundary of a domain
where we can obtain a contribution from the
retarded Green's function integrated over sources.

\section{Electromagnetic fields:
the retarded solutions}
\label{sc:EMRetField}

In this section we shall analyze the electromagnetic
fields of free or accelerated charges
with monopole and also with a dipole structure.
We shall pay attention to the constraints
which the electromagnetic field, in contrast to the
scalar field, has to satisfy on any spacelike
hypersurface.

\subsection{Free monopole}
\label{ssc:FreeMon}

Let us start with an unaccelerated monopole at rest
at the origin of both coordinate systems used, i.e.,
at $\tlr=R=0$. With $R_\oix=0$, the potential
\eqref{EMAsym-amon} and the field \eqref{EMFsymRT-amon}
simplify to
\begin{equation}\label{EMAsym-mon}
\EMA_\sym = - \frac{\EMq}{4\pi}
\frac{\sin\tlt}{\cos\tlt+\cos\tlr}
\Bigl(\frac{1+\cos\tlt\cos\tlr}{\sin\tlt\sin\tlr}\,
\grad\tlt + \grad\tlr\Bigr)
\end{equation}
and
\begin{equation}\label{EMFsym-mon}
\EMF_\sym = - \frac{\EMq}{4\pi}
\frac{1}{\sin^2\tlr}\grad\tlt\wedge\grad\tlr\period
\end{equation}

Let us restrict the potential to the \vague{creation
light cone} and its interior by defining
\begin{equation}\label{EMAo-mon}
\EMA_\cut = \EMA_\sym\;\stepfc(\tlu)\period
\end{equation}
The field in null coordinates $\tlu$, $\tlv$ then reads
\begin{equation}\label{EMFo-mon}
\EMF_\cut = \grad\EMA_\cut =
\EMF_\sym\;\stepfc(\tlu) -
\frac{\EMq}{4\pi}\frac{1}{\sin\tlv}\;
\deltafc(\tlu)\;\grad\tlu\wedge\grad\tlv\commae
\end{equation}
so that the left-hand side of Maxwell's equations becomes
\begin{equation}\label{MEo-mon}
\begin{split}
\stcnx_{\mu}\EMF_\cut^{\alpha\mu} & \;=\; \EMJ_\mon^\alpha \\
& + \frac{\EMq}{4\pi\alpha^4} \Bigl(
\frac{1-\cos\tlv}{1+\cos\tlv}\cvect[\alpha]{\tlu} +
2 (1-\cos\tlv)\cvect[\alpha]{\tlv} \Bigr)\; \deltafc(\tlu) \\
& - \frac{\EMq}{4\pi\alpha^4} \cvect[\alpha]{\tlv}\;
\deltafc^\prime(\tlu) \period
\end{split}
\end{equation}
Here $\EMJ_\mon^\alpha = (\stcnx_{\mu}\EMF_\sym^{\alpha\mu})
\stepfc(\tlu)$ is the current produced by the charge at
$\tlr=0$. Additional terms on the right-hand side of
Eq.~\eqref{MEo-mon}, localized on the null hypersurface
$\tlu=0$, clearly show that the restricted field
\eqref{EMAo-mon} does not correspond to a single point
source. The terms of this type did not arise in case
of the scalar field discussed in the previous section.

We can try to add a field localized on $\tlu=0$,
which would cancel the additional terms. Although we
shall see in the following subsection that this cannot
be achieved, it is instructive to add, for example,
the field
\begin{equation}\label{EMFc-mon}
\EMA_\cmp = - \frac{\EMq}{4\pi}\ln\Bigl(\tan\frac{\tlv}{2}\Bigr)\,
\deltafc(\tlu)\;\grad\tlu\commae
\end{equation}
which cancels the second term on the right hand side of the
field \eqref{EMFo-mon}. Thus denoting,
\begin{equation}\label{EMmon}
\begin{aligned}
\EMA_\mon &= \EMA_\cut + \EMA_\cmp\commae\\
\EMF_\mon &= \grad\EMA_\mon =
\EMF_\sym\;\stepfc(\tlu)\commae
\end{aligned}
\end{equation}
we find that with $\EMF_\mon$, Maxwell's equations
become
\begin{equation}\label{MEmon}
\stcnx_\mu\EMF_\mon^{\alpha\mu} =
\EMJ_\mon^\alpha
- \frac{\EMq}{4\pi\alpha^4} (1-\cos\tlv)\;
\deltafc(\tlu)\;\cvect[\alpha]{\tlv}\period
\end{equation}
Hence, the field $\EMA_\mon$ does not represent only
the unaccelerated monopole charge but also
a spherical shell of charges moving outwards from
the monopole with the velocity of light along
the \vague{creation light cone} $\tlu=0$.
The total charge of the shell is precisely opposite
to the monopole charge so that the total
charge of the system is zero.

We shall return to this point in the following subsection,
now let us add yet two comments. It is interesting that,
in contrast to the scalar field strength
\eqref{dSF-amon}, the electromagnetic field
$\EMF_\mon$ is not singular at $\tlu=0$.
Apparently, the effects of the monopole and
the charged shell compensate along $\tlu=0$
in such a way that even the energy-momentum
tensor of the field is finite there.

Second, if the field $\EMA_\mon$, corresponding
to the retarded field from the charge $\EMq$ at
$\tlr=0$ and the outgoing charged shell is
superposed with the analogous field
corresponding to the advanced field from
the charge $-\EMq$ at $\tlr=\pi$ and the
ingoing charged shell, the fields
corresponding to charged shells localized on
$\tlu=0$ cancel each other and the field
$\EMA_\sym$ (Eq.~\eqref{EMAsym-amon}) with
$R_\oix=0$ is obtained. The same compensation occurs
for two uniformly accelerated charges ($R_\oix\ne 0$)
considered in Section~\ref{sc:SymField},
as it follows from the symmetry. Therefore,
the symmetric electromagnetic field
\eqref{EMAsym-amon} can be interpreted as
arising from the combinations of retarded and
advanced potentials due to both charges
$1$ and $1'$ in the same way as was the case for
the symmetric scalar field; relation
\eqref{symRAfield} remains true if $\SF$'s are
replaced by $\EMA$'s.

\subsection{Constraints}
\label{ssc:EMConstr}

The appearence of a shell with the total charge
exactly opposite to that of the monopole discussed
above has deeper reasons. It is a consequence
of the constraints, which any electromagnetic field,
and charge distributions have to satisfy on
a spatial hypersurface and of the fact
that spatial hypersurfaces, including past and
future infinities, in de~Sitter spacetime,
are compact. Integrating the constraint equation
\begin{equation}\label{Gauss}
\scnx_\mu\EME^\mu = \EMr
\end{equation}
(see Eq.~\eqref{EMEdef} for the definition of $\EME^\alpha$)
over a compact Cauchy hypersurface $\Sigma$, we convert
the integral of the divergence on the
left-hand side to the integral over a
\vague{boundary} which, however, does not exist for
a compact $\Sigma$. Hence, as it is well known, the
total charge on a compact hypersurface
(in any spacetime, not only de~Sitter)
must vanish:
\begin{equation}\label{TotChargeConstr}
Q_\tot = 0\period
\end{equation}
Therefore, the field $\EMF_\mon$ constructed
in \eqref{EMmon} represents the monopole field plus the
\vague{simplest} additional source localized on
the past horizon of the monopole that leads
to the total zero charge. This enables
the monopole electric field lines to end on
this horizon.

A stronger, even \emph{local} condition on the
charge distribution in de~Sitter spacetime
(or, indeed, in any spacetime with spacelike
past infinity $\scry^-$) arises if we admit
purely retarded fields only. Here we
define \defterm{purely retarded fields} as those
that \emph{vanish at $\scry^-$}. Then, however,
the constraint \eqref{Gauss} directly
implies that at $\scry^-$ the
charge distribution vanishes:
\begin{equation}\label{LocChargeConstr}
\EMr|_{\tlt=0} = 0\period
\end{equation}

In the Les~Houches lectures in 1963 Penrose
\cite{Penrose:1964} gave a general argument
showing that if $\scry^-$ is spacelike and
the charges meet it in a discrete set of
points, then there will be incosistencies
if an incoming field is absent (see in particular
Fig.~16 in Ref.~\cite{Penrose:1964},
cf. Fig.~\ref{fig:Penrose},
see also Ref.~\cite{Penrose:1967}). Penrose also
remarked that an alternative definition
of advanced and retarded
fields might be found that leads to
different results, and that the application
of the result to physical models is not clear.
We found nothing more
on this problem in the literature
since Penrose's observation in 1963.
Our work appears to give the first
explicit model in which this issue can
be analyzed.

\subsection{Rigid dipole}
\label{ssc:RigDip}

As the first example of a simple source satisfying
both the constraint \eqref{TotChargeConstr}
and the local condition \eqref{LocChargeConstr}
required by the absence of incoming radiation,
we consider a rigid dipole. To construct an
elementary rigid dipole, we place point charges
${\EMq}/{\epsilon}$ and $-{\EMq}/{\epsilon}$
on the worldlines with $R_\oix=\frac12{\epsilon\alpha}$
and $\vartheta=0,\pi$,
fixed in the static coordinates,
and take the limit $\epsilon\rightarrow 0$.
The constant dipole moment is thus given by
$\EMd=\EMq \alpha$. The resulting symmetric field
can easily be deduced from the symmetric fields
\eqref{EMAsym-amon}--\eqref{EMFsymtl-amon} of
electric monopoles:\footnote{
  Here in the potential we ignore a trivial gauge term
  proportional to $\grad\cos\vartheta$.
  }
\begin{equation}\label{EMsym-rdip}
\begin{aligned}
\EMA_\sym &=
- \frac{\EMd}{4\pi\alpha}
\frac{\cos\vartheta}{\sin^2\tlr}\,
\Bigl(\sin\tlt\cos\tlr\,\grad\tlt -
\cos\tlt\sin\tlr\,\grad\tlr\Bigr)\commae\\
\EMF_\sym &= - \frac{\EMd}{4\pi\alpha}\,
\Bigl( 2\cos\vartheta\frac{\sin\tlt}{\sin^3\tlr}\,
\grad\tlt\wedge\grad\tlr \\
&\qquad\qquad-\frac{\sin\vartheta}{\sin^2\tlr}\bigl(
\sin\tlt\cos\tlr\,\grad\vartheta\wedge\grad\tlt -
\cos\tlt\sin\tlr\,\grad\vartheta\wedge\grad\tlr
\bigr)\Bigr)\period
\end{aligned}
\end{equation}
As in Section~\ref{sc:SymField}, $R_\oix=0$ corresponds
to two worldlines and the symmetric solution
\eqref{EMsym-rdip} describes the fields of
\emph{two} dipoles, one at rest at $\tlr=0$,
the other at $\tlr=\pi$.

To construct a purely retarded field of the
dipole at $\tlr=0$, we first restrict the symmetric
field \eqref{EMsym-rdip} to the inside of the
\vague{creation light cone} of the dipole,
analogously as we did with the monopole charge in
Subsection~\ref{sc:EMRetField}.\ref{ssc:FreeMon}. Writing
(cf.~\eqref{EMAo-mon})
\begin{equation}\label{EMAo-rdip}
\EMA_\cut = \EMA_\sym\;\stepfc(\tlu)\commae
\end{equation}
we now get
\begin{equation}\label{EMFo-rdip}
\EMF_\cut = \EMF_\sym\;\stepfc(\tlu)\commae
\end{equation}
so that no additional term like that in
Eq.~\eqref{EMFo-mon} arises; however, expressing
the left-hand side of Maxwell's equations as
in Eq.~\eqref{MEo-mon}, we find
\begin{equation}\label{MEo-rdip}
\stcnx_\mu\EMF_\cut^{\alpha\mu} = \EMJ_\rdip^\alpha -
\frac{\EMd}{4\pi\alpha}\,2\cos\vartheta\,
(1-\cos\tlv)\,\deltafc(\tlu)\;
\cvect[\alpha]{\tlv}\period
\end{equation}
Here  $\EMJ_\rdip^\alpha =
(\stcnx_\mu\EMF_\sym^{\alpha\mu})\stepfc(\tlu)$
is the current corresponding to the rigid dipole at
$\tlr=0$. Similarly to the case of the monopole,
there is an additional term on the right-hand side of
Eq.~\eqref{MEo-rdip}, localized on the null cone
$\tlu=0$, indicating that the field
\eqref{EMAo-rdip} represents, in addition
to the dipole, an additional source located
on the horizon $\tlu=0$. In contrast
to the monopole case, however, this source can
be compensated by adding to the potential
\eqref{EMAo-rdip} the term
\begin{equation}\label{EMAcmp-rdip}
\EMA_\cmp = - \frac{\EMd}{4\pi\alpha}\cos\vartheta\;
\deltafc(\tlu)\;\grad\tlu\period
\end{equation}
In this way we finally obtain the purely
retarded field of the dipole with dipole
moment $\EMd$, located at $\tlr=R=0$, in the form
\begin{equation}\label{EM-rdip}
\begin{aligned}
\EMA_\rdip &= \EMA_\cut+\EMA_\cmp\commae\\
\EMF_\rdip &= \EMF_\sym\;\stepfc(\tlu)
+ \frac{\EMd}{4\pi\alpha}
\sin\vartheta\;\deltafc(\tlu)\;
\grad\vartheta\wedge\grad\tlu\commae
\end{aligned}
\end{equation}
where $\EMA_\cut$, $\EMA_\cmp$, $\EMF_\sym$ are
given by Eqs.~\eqref{EMAo-rdip}, \eqref{EMAcmp-rdip},
and \eqref{EMsym-rdip}. It is easy to check
that $\stcnx_\mu\EMF_\rdip^{\alpha\mu}=
\EMJ_\rdip^\alpha$ is satisfied.

Regarding the retarded field \eqref{EM-rdip},
we see that it is, in contrast to the symmetric
field \eqref{EMsym-rdip}, singular on the
\vague{creation light cone} (past event horizon)
$\tlu=0$. This is not surprising; in order
to obtain a purely retarded field, we \vague{squeezed}
the field lines produced by the dipole into the horizon.

\subsection{Geodesic dipole}
\label{ssc:GeodDip}


Next we consider dipoles consisting of two free
charges moving along the geo\-desic\footnote{
  Charges are called free in the sense that
  they are assumed to be moving along geodesics.
  Of course, there is an electromagnetic
  interaction between them, which is neglected
  or has to be compensated.}
$r=\text{constant}$,
$\vartheta=\text{constant}$, $\varphi=\text{constant}$
in the Minkowski space which, as discussed at the end of
Section~\ref{sc:PointPart} (see Eq.~\eqref{RigMon} and
the worldlines $2$, $2'$ in Fig.~\ref{fig:Worldlines})
transforms into geodesics of the conformally
related de~Sitter space. We call two free opposite
charges a \defterm{geodesic dipole}.

We start again by constructing first the symmetric
field. Two elementary geodesic dipoles located
at $\tlr=0$ and $\tlr=\pi$ can be obtained by
placing point charges $\pm\EMq/{\epsilon}$
on the worldlines $r=\epsilon\alpha/2$,
$\vartheta=0,\pi$ and taking the limit
$\epsilon\rightarrow 0$. As with monopoles,
to find the symmetric field we conformally transform
the field of a standard rigid Minkowski dipole
(bewaring the signs for $t>0$ and $t<0$ so that the
field is analytic outside the sources). The dipole
moment in Minkowski space $\EMf=\EMq\alpha$ is
constant, the corresponding value in de~Sitter spacetime,
however, depends now on time:
\begin{equation}\label{EMd-gdip}
\EMd=\frac{\alpha}{t}\EMf=
\frac{\pm 1+\cos\tlt}{\sin\tlt}\,\EMq\alpha\commae
\end{equation}
as it follows from the transformation relations
\eqref{EMrCT} and \eqref{trTtlttlr}. In terms of $\EMf$ we find
the symmetric field of geodesic dipoles located
at $\tlr=0$ and $\tlr=\pi$ to read
\begin{equation}\label{EMsym-gdip}\raisetag{42pt}
\begin{aligned}
\EMA_\sym &= - \frac{\EMf}{4\pi}\,
\frac{\cos\vartheta}{r^2}\, \grad t =
- \frac{\EMf}{4\pi\alpha^2}
\frac{\cos\vartheta}{\sin^2\tlr}\,
\Bigl( (1+\cos\tlt\cos\tlr)\,\grad\tlt +
\sin\tlt\sin\tlr\;\grad\tlr \Bigr)\commae\\
\EMF_\sym &= - \frac{\EMf}{4\pi\alpha^2}\,
\Bigl( 2\frac{\cos\vartheta}{\sin^3\tlr}\,
(\cos\tlt+\cos\tlr)\,\grad\tlt\wedge\grad\tlr \\
&\qquad-\frac{\sin\vartheta}{\sin^2\tlr}\bigl(
(1+\cos\tlt\cos\tlr)\,\grad\vartheta\wedge\grad\tlt +
\sin\tlt\sin\tlr\,\grad\vartheta\wedge\grad\tlr
\bigr)\Bigr)\period
\end{aligned}
\end{equation}
Proceeding as with the rigid dipole, we
generate the retarded field of only one
geodesic dipole at $\tlr=0$ by restricting
the symmetric field by the step function:
\begin{equation}\label{EMo-gdip}
\begin{aligned}
\EMA_\cut &= \EMA_\sym\,\stepfc(\tlu)\commae\\
\EMF_\cut &= \EMF_\sym\,\stepfc(\tlu)
- \frac{2\EMf}{4\pi\alpha^2}
\frac{\cos\vartheta}{1-\cos\tlv}\,\deltafc(\tlu)\;
\grad\tlv\wedge\grad\tlu\period
\end{aligned}
\end{equation}
Since
\begin{equation}\label{MEo-gdip}
\begin{split}
\stcnx_\mu\EMF_\cut^{\alpha\mu} &= \EMJ_\gdip^\alpha
+\frac{2\EMf}{4\pi\alpha^6}\,
\deltafc(\tlu)\,\Bigl(
-\sin\vartheta\,\cvect[\alpha]{\vartheta}+
\cos\vartheta\sin\tlv\,\cvect[\alpha]{\tlv}\Bigr)\\
&\qquad\qquad-\frac{2 \EMf}{4\pi\alpha^6}\,
\cos\vartheta\,(1-\cos\tlv)\;
\deltafc^\prime(\tlu)\,\cvect[\alpha]{\tlv}\commae
\end{split}
\end{equation}
where $\EMJ_\gdip^\alpha =
(\stcnx_\mu\EMF_\sym^{\alpha\mu})\stepfc(\tlu)$,
an additional source is present at the horizon
$\tlu=0$; it can be compensated by adding an
additional field with potential,
for example, given by
\begin{equation}
\EMA_\cmp = -\frac{2\EMf}{4\pi\alpha^2}
\;\deltafc(\tlu)\,\Bigl(
\cos\vartheta\frac{\sin\tlv}{1-\cos\tlv}\,\grad\tlu
+\sin\vartheta\,\grad\vartheta\Bigr)\period
\end{equation}
The total retarded field of the geodesic
dipole is then given as follows:
\begin{equation}\label{EM-fdip}
\begin{aligned}
\EMA_\gdip &= \EMA_\cut + \EMA_\cmp\commae\\
\EMF_\gdip &= \EMF_\sym\,\stepfc(\tlu)\\
&\quad+ \frac{2\EMf}{4\pi\alpha^2}
\;\deltafc(\tlu)\,\Bigl(
\frac{2\cos\vartheta}{1-\cos\tlv}\,
\grad\tlv\wedge\grad\tlu +
\sin\vartheta\frac{\sin\tlv}{1-\cos\tlv}\,
\grad\vartheta\wedge\grad\tlu\Bigr)\\
&\quad- \frac{2\EMf}{4\pi\alpha^2}\,\sin\vartheta\,
\deltafc^\prime(\tlu)\;\grad\vartheta\wedge\grad\tlu\period
\end{aligned}
\end{equation}
It indeed satisfies Maxwell's equations:
\begin{equation}
\stcnx_\mu\EMF_\gdip^{\alpha\mu} = \EMJ_\gdip^\alpha\period
\end{equation}

\section{Conclusion}
\label{sc:summary}

\enlargethispage*{2em}

By using the conformal relation between de Sitter and Minkowski
space, we constructed various types of fields produced by scalar
and electromagnetic charges moving with a uniform (possibly zero)
acceleration in de Sitter background. One of our main conclusions
has been the explicit confirmation and elucidation of the
observation by Penrose that in spacetimes with a spacelike past
infinity, which implies the existence of a particle horizon,
purely retarded fields do not exist for general source
distributions.

In Subsection~\ref{sc:EMRetField}.\ref{ssc:GeodDip}
we constructed the field \eqref{EM-fdip} of the
geodesic dipole which, at first sight, appears as being a
purely retarded field. Can thus purely retarded fields of, for
example, two opposite monopoles, be constructed by distributing the
elementary geodesic dipoles along a segment $\tlr\in\langle 0,a\rangle$,
$\vartheta = \varphi = 0$, so that neighboring opposite
charges cancel out, and only two monopoles, at $\tlr=0$
and $\tlr = a$, remain? Then, however, the local
constraint \eqref{LocChargeConstr} on the charge distribution
would clearly be violated!

The solution of this paradox is in the fact that the field \eqref{EM-fdip}
\emph{is not a purely} retarded field; if we calculate (by using distributions)
the initial data leading to the field
strength \eqref{EM-fdip}, we discover that
the electric field strength contains terms proportional
to the $\deltafc$-function at $\tlt=0$, $\tlr=0$ so that
the field does not vanish at $\scry^-$.
Hence the field of the dipoles distributed along the segment does
\emph{not} vanish at $\scry^-$ and thus cannot be considered as
a purely retarded field of the two monopoles at $\tlr=0$
and $\tlr=a$.

We thus arrive at the conclusion that a purely retarded field of
even two opposite charges (so that the global constraint of a
zero net charge \emph{is} satisfied) cannot be constructed in
the de Sitter spacetime unless the charges \vague{enter} the universe at
the same point at $\scry^-$ and the \emph{local} constraint
\eqref{LocChargeConstr} is satisfied. If we allow
nonvanishing initial data at $\scry^-$, the resulting
fields can hardly be considered as \vague{purely retarded.}

By applying the superposition principle we can
consider a greater number of sources, and our
arguments of the insufficiency of purely retarded
fields (based on the global and local constraints)
can clearly be generalized also to infinitely
many discrete sources. Our discussion of the fields of dipoles
indicates that interesting situations may arise.

The absence of purely retarded fields in de Sitter spacetime
or, in fact, in any spacetime with space-like $\scry^-$
is, of course, to be expected to occur for higher-spin fields as well.
In particular, there has been much interest in the primordial
gravitational radiation. Since de Sitter spacetime is a standard
arena for inflationary models, the generation of gravitational
waves by (test) sources in de Sitter spacetime has been studied in
recent literature (see Ref.~\cite{VegaRamirezSanchez:1999},
and references therein).  We plan to analyze
the linearized gravity on de Sitter background in
the light of the results described above. The fact that particles
are expected to have only positive gravitational mass will
apparently prevent any purely retarded field to exist.

First, however, we shall present the generalization
of the well known Born's solution for uniformly accelerated charges
in Minkowski spacetime to the generalized
Born solution representing uniformly
accelerated charges in de Sitter spacetime
\cite{BicakKrtous:BIS}. As it was demonstrated
in the present-paper, to be \vague{born in de Sitter} is quite a
different matter than to be \vague{born in Minkowski}.

\section*{Acknowledgments}

The authors would like to thank the Albert-Einstein  Institute, Golm, where
part of this work was done, for its kind hospitality. J.B. also
acknowledges the Sackler Fellowship at the Institute of Astronomy in Cambridge,
where this work was finished. We were supported in part by
Grants Nos. GA\v{C}R 202/99/026 and GAUK 141/2000 of the Czech
Republic and Charles University.

\appendix

\section{Coordinate systems in de~Sitter space}
\label{apx:Coor}

We describe here coordinate systems employed in the main text.
There exists extensive literature on various useful coordinates
in de~Sitter spacetime --- for standard reference see
Ref.~\cite{HawkingEllis:book}, for more recent reviews,
containing also many references, see, for example,
Refs.~\cite{Schmidt:1993,EriksenGron:1995}.

{\nohyphenationafterdash
de~Sitter spacetime has topology $S^3\times\realn$.
It is best visualized as the four-di\-men\-sional
hyperboloid imbedded in flat five-di\-men\-sional Minkowski
space; it is the homogeneous space
of constant curvature spherically
symmetric about any point. The following coordinate
systems are constructed around any fixed (though arbitrary)
point. Two of the coordinates are just standard
spherical angular coordinates $\vartheta\in\langle0,\pi\rangle$
and $\varphi\in\langle-\pi,\pi\rangle$ on the orbits of the
Killing vectors of spherical symmetry with the homogeneous metric
which is
\begin{equation}
\sphmtrc = \grad\vartheta\formsq+
\sin^2\theta\;\grad\varphi\formsq\period
\end{equation}
In the transformations considered below, these coordinates
remain unchanged and thus are not written down.
The axis is fixed by $\vartheta=0,\pi$.
}

Next, we have to introduce time and radial coordinates
labeling the orbits of spherical symmetry.
The coordinate systems defined below
differ only in these time and radial
coordinates and, therefore, we essentially work
with a two-dimensional system.
Radial coordinates commonly take
positive values and coordinate systems
are degenerate for their value zero.

As discussed in Section~\ref{sc:MinkDSCompact}
below Eq.~\eqref{CFfact}, the relations
between various regions I--IV of Minkowski and
de~Sitter spaces are conveniently described
if we allow radial coordinates to attain negative
values. We adopt the convention that at a fixed time
the points symmetrical with respect to the origin
of spherical coordinates have the opposite sign of
the radial coordinate. Hence, the points with
$\{t,r,\vartheta,\varphi\}$ are identical with
$\{t,-r,\pi-\vartheta,\varphi+\pi\}$, analogously
for the coordinates $\{\tlt,\tlr,\vartheta,\varphi\}$
introduced below.

Let us characterize this convention in more detail.
Imagine that on our manifold (either compactified
Minkowski space or de~Sitter space) two coordinate
maps $\{t_-,r_-,\vartheta_-,\varphi_-\}$ and
$\{t_+,r_+,\vartheta_+,\varphi_+\}$ are introduced,
which for any fixed point are connected by the relations
\begin{equation}
\begin{gathered}
t_-=t_+\comma
r_-=-r_+\commae\\
\vartheta_-=\pi-\vartheta_+\comma
\varphi_-=\varphi_++\pi\!\!\mod2\pi\commae
\end{gathered}
\end{equation}
where $r_+\in\realn^+$ and $r_-\in\realn^-$.
\emph{Both} maps cover the \emph{whole} spacetime manifold.
Now we consider a two-dimensional cut given by
$\vartheta_+=\vartheta_\oix$, $\varphi_+=\varphi_\oix$,
with $t_+$, $r_+$ changing. This represents
the history of a half-line $l_+$ in
Fig.~\ref{fig:Mink}, illustrated by regions I
and III, say. The history of a half-line
$l_-$, obtained by the smooth extension of
$l_+$ through the origin, illustrated by
regions II and IV, is covered by
the coordinates $\{t_-,r_-,\vartheta_-,\varphi_-\}$
with the \emph{same} angular coordinates
${\vartheta_-=\vartheta_\oix}$, ${\varphi_-=\varphi_\oix}$
but $r_-<0$ (which, in our convention, is
identical to ${r_+>0}$,
${\vartheta_+=\pi-\vartheta_\oix}$,
${\varphi_+=\varphi_\oix+\pi}$).

In exactly the same way we may introduce coordinates
$\{\tlt_\pm,\tlr_\pm,\vartheta,\varphi\}$,
with $\{\tlt_+,\tlr_+\}$ covering regions I and IV
in Fig.~\ref{fig:dS}, whereas $\{\tlt_-,\tlr_-\}$ cover
regions II and III. In Fig.~\ref{fig:Coord}, the
two-dimensional cuts (with angular coordinates fixed)
through de~Sitter space
(or through the compactified space $M^\#$)
are illustrated. Both regions covered by
$\{\tlt_+,\tlr_+\}$ and by $\{\tlt_-,\tlr_-\}$
are included (the right and left parts of the diagrams).
However, in the figures, as well as in
the main text, we do not write down subscripts
``$+$'' and ``$-$'' at the coordinates, since the
sign of the radial coordinate specifies which
map is used.

de~Sitter spacetime can be covered by
\defterm{standard} coordinates $\tau$, $\chi$
($\tau\in\realn$, $\chi\in\realn^+$)
in which the metric has the form \cite{HawkingEllis:book}:
\begin{equation}
\mtrc = -\grad\tau\formsq+
\alpha^2\cosh^2\!\frac{\tau}{\alpha}\;\Bigl(
\grad\chi\formsq + \sin^2\chi\;\sphmtrc\Bigr)\period
\end{equation}
It is useful to rescale the coordinate $\tau$ to
obtain \defterm{conformally Einstein} coordinates
$\{\tlt, \tlr\}$:
\begin{gather}
\begin{aligned}
\tlt &= 2 \arctan\Bigl(\exp{\frac{\tau}{\alpha}}\Bigr)
\comma&\tlt&\in\langle0,\pi\rangle\commae\\
\tlr &= \chi
\comma&\tlr&\in\langle0,\pi\rangle\commae
\end{aligned}\label{dScoorHom}\\
\mtrc = \frac{\alpha^2}{\sin^2\tlt}
\bigl(-\grad\tlt\formsq+\grad\tlr\formsq
+\sin^2\tlr\;\sphmtrc\bigr)\period\label{dSmtrcHom}
\end{gather}
In these coordinates, de~Sitter space is explicitly
seen to be conformal to the part of the Einstein
static universe. Coordinate lines are drawn in
Fig.~\ref{fig:Coord}(b).

Another coordinate system used in our work
are inertial coordinates $t$, $r$ of conformally
related Minkowski space. We call them
\defterm{conformally flat} coordinates:
\begin{gather}
\begin{aligned}
t &= \frac{\alpha\sin\tlt}
{\cos\tlr+\cos\tlt}\comma&
\tlt &= \arctan\frac{2\,t\,\alpha}
{\alpha^2-t^2+r^2}
\comma&t&\in\realn\commae\\
r &= \frac{\alpha\sin\tlr}
{\cos\tlr+\cos\tlt}\comma&
\tlr &= \arctan\frac{2\,r\,\alpha}
{\alpha^2+t^2-r^2}
\comma&r&\in\realn^+\commae
\end{aligned}\label{dScoorConf}\\
\mtrc = \frac{\alpha^2}{t^2}
\Bigl(-\grad t\formsq+\grad r\formsq
+r^2\;\sphmtrc\Bigr)\period\label{dSmtrcConf}
\end{gather}
Coordinate lines are drawn in Fig.~\ref{fig:Coord}(c).

\begin{figure*}[t]
  \begin{center}
  \epsfig{file=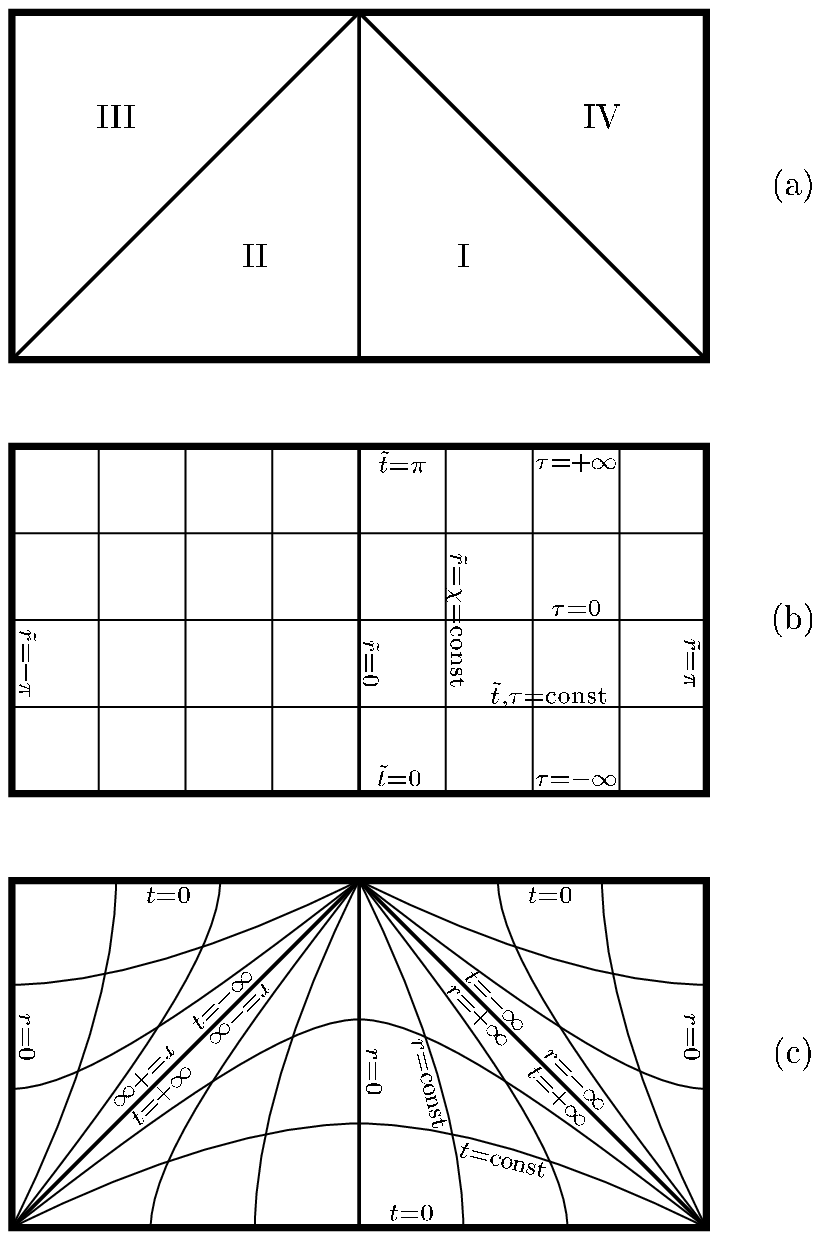,width=0.74\textwidth}
  \end{center}
\end{figure*}
\begin{figure}[t]
  \begin{center}
  \epsfig{file=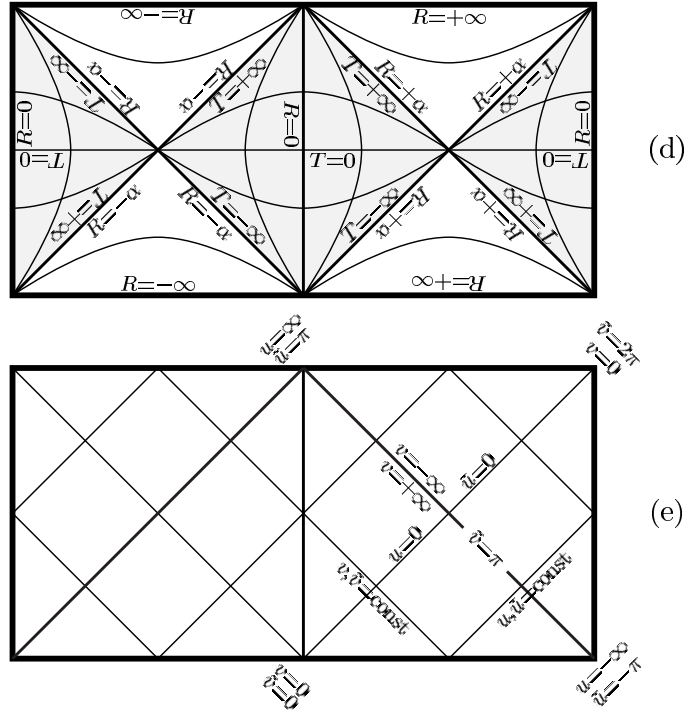,width=0.74\textwidth}
  \end{center}
  \caption{Coordinates in de~Sitter space}
  \figdescription{
   (a) Regions I--IV are specified.
   They correspond to the same
   regions as in the cut $\mathcal{C}$ in
   Fig.~\ref{fig:dS}.
   Coordinate lines of the conformally
   Einstein coordinates (b), of the conformally flat
   coordinates (c), of the static coordinates (d), and
   of the null coordinates (e), are indicated.
   For the definition of all these coordinate
   systems, see Eqs.~\eqref{dScoorHom}--\eqref{dSmtrcNull}.
   All the figures describe
   the same cut of de~Sitter space.
   The ranges of coordinates covering the cut,
   as well as directions in which they grow,
   can be seen from the figures.
   }
  \label{fig:Coord}
\end{figure}

Commonly used are \defterm{static} coordinates
$T$, $R$, related to the time-like
Killing vector $\cvect{T}$ of de~Sitter spacetime:
\begin{gather}
\begin{aligned}
T &= \frac\alpha2\,
\log\frac{t^2-r^2}{\alpha^2} =
\frac\alpha2\,
\log\frac{\cos\tlr-\cos\tlt}
{\cos\tlr+\cos\tlt}
\comma&T&\in\realn\commae\\
R &= \alpha\;\frac{r}{t} =
\alpha\,\frac{\sin\tlr}{\sin\tlt}
\comma&R&\in\langle0,\alpha\rangle\commae
\end{aligned}\label{dScoorStat}\\
\mtrc = -\Bigl(1-\frac{R^2}{\alpha^2}\Bigr)
\grad T\formsq +
\Bigl(1-\frac{R^2}{\alpha^2}\Bigr)^{\!\!-1}
\grad R\formsq + R^2 \sphmtrc\period\label{dSmtrcStat}
\end{gather}
As it is well-known, these coordinates do not cover the whole
spacetime but only the domain with $\tlt+\tlr<\pi$
and $\tlt-\tlr>0$. The boundary of this
domain is the Killing horizon.
The coordinate $R$ can be extended
smoothly to the whole spacetime but it
is not unique globally.
It is also useful to rescale coordinate $R$
to obtain the \defterm{expanded static}
coordinates $\brt$, $\brr$:
\begin{gather}
\begin{aligned}
\brt &= T
\comma&\brt&\in\realn\commae\\
\brr &= \alpha\;\arctanh\frac{R}{\alpha}
\comma&\brr&\in\realn\commae
\end{aligned}\label{dScoorEStat}\\
\mtrc = \Bigl(\cosh\frac{\brr}{\alpha}\Bigr)^{-2}\;
\Bigl(-\grad\brt\formsq + \grad\brr\formsq
+\alpha^2\sinh^2\!\frac{\brr}{\alpha}\;\,\sphmtrc\Bigr)\period\label{dSmtrcEStat}
\end{gather}
Coordinate lines for the static coordinates
are drawn in Fig.~\ref{fig:Coord}(d).

Finally, three sets of \defterm{null}
coordinates $\{u,v\}$, $\{\bru,\brv\}$, and $\{\tlu,\tlv\}$
are defined by
\begin{equation}\label{dScoorNull}
\begin{aligned}
u &= t-r \comma& \tlu &= \tlt-\tlr \comma&
\bru &= \brt-\brr \commae\\
v &= t+r \comma& \tlv &= \tlt+\tlr \comma&
\brv &= \brt+\brr \period
\end{aligned}
\end{equation}
From here we find
\begin{equation}
u = \alpha\,\tan\frac\tlu2 =
\alpha\,\exp{\frac\bru\alpha}\comma
v = \alpha\,\tan\frac\tlv2 =
\alpha\,\exp{\frac\brv\alpha}\period
\end{equation}
The metric in these coordinates reads
\begin{equation}\label{dSmtrcNull}
\begin{aligned}
\mtrc
&= \frac{\alpha^2}{(u+v)^2}\;
\Bigl(-2\,\grad u \vee \grad v +
(u-v)^2\,\sphmtrc\Bigr) \\
&= \frac{\alpha^2}{1-\cos(\tlv+\tlu)}\;
\Bigl(-\,\grad \tlv \vee \grad \tlu +
\bigl(1-\cos(\tlv-\tlu)\bigr)\,\sphmtrc\Bigr) \\
&= \bigl(e^{\bru/\alpha}
+e^{\brv/\alpha}\bigr)^{\!-2}\;
\Bigl(-2\,e^{(\bru+\brv)/\alpha}\;
\grad \bru \vee \grad \brv +
\alpha^2\,\bigl(e^{\bru/\alpha}
-e^{\brv/\alpha}\bigr)^2\,\sphmtrc\Bigr)
\period
\end{aligned}
\end{equation}
The corresponding coordinate lines are illustrated in
Fig.~\ref{fig:Coord}(e).



\end{document}